\documentclass[final,5p,times,authoryear]{elsarticle}

\usepackage{amssymb}
\usepackage{amsmath}
\usepackage{tikz}
\graphicspath{{./}{figures/}}
\usetikzlibrary{decorations.pathreplacing}
\usepackage{xcolor}
\definecolor{xlinkcolor}{cmyk}{1.0, 0.31, 0, 0.42}

\PassOptionsToPackage{hyphens}{url}
\usepackage[bookmarks=true, pdfnewwindow=true, colorlinks=true, linkcolor=xlinkcolor, citecolor=xlinkcolor, filecolor=xlinkcolor, urlcolor=xlinkcolor, final=true]{hyperref}

\journal{New Astronomy}

\begin{document}
\begin{frontmatter}



\title{Multi-Point Hermite Methods for the $N$-body Problem}


\author[inst1,inst2,label1]{Alexander J. Dittmann}
\fntext[label1]{NASA Einstein Fellow}
\affiliation[inst1]{organization={Institute for Advanced Study},
            addressline={1 Einstein Drive}, 
            city={Princeton},
            postcode={08540}, 
            state={NJ},
            country={USA}}


\affiliation[inst2]{organization={Department of Astronomy and Joint Space-Science Institute},
            addressline={University of Maryland}, 
            city={College Park},
            postcode={20742}, 
            state={MD},
            country={USA}}
\ead{dittmann@ias.edu}
\begin{abstract}
Numerical integration methods are central to the study of self-gravitating systems, especially those comprised of many bodies or otherwise beyond the reach of analytical methods. Predictor-corrector schemes, both multi-step methods and those based on 2-point Hermite interpolation, have found great success in the simulation of star clusters and other collisional systems. Higher-order methods, such as those based on Gaussian quadratures and Richardson extrapolation, have also proven popular for high-accuracy integrations of few-body systems, particularly those that may undergo close encounters. This work presents a family of high-order schemes based on multi-point Hermite interpolation. When applied as multi-step multi-derivative schemes, these can be seen as generalizing both Adams-Bashforth-Moulton methods and 2-point Hermite methods; I present results for the 6th-, 9th-, and 12th-order 3-point schemes applied in this manner using variable time steps. In a star cluster-like test problem, the 3-point 6th-order predictor-corrector scheme matches or outperforms the standard 2-point 4th-order Hermite scheme at negligible $\mathcal{O}(N)$ additional cost, potentially reducing the necessary number of force evaluations in simulations of large-$N$ collisional systems by factors of $\sim 3$ or more. I also present a number of high-order time-symmetric schemes up to 18th order, which have the potential to improve the accuracy and efficiency of long-duration simulations.
\end{abstract}




\end{frontmatter}

\section{Introduction}\label{sec:intro}
The motions of stars and planets have been studied mathematically since antiquity \citep[see, for example,][]{neugebauer1969exact}; their study was revolutionized by the work of Kepler and his penecontemporaries \citep[e.g.,][]{kepler1609nova, kepler1619harmonices, 1687pnpm.book.....N}, and was again transformed following the advent of electronic computers \citep[e.g.][]{1963ZA.....57...47V,1998sssc.book.....M,2007NewA...12..641P}.\footnote{Notably, mechanical calculations with excellent ($\mathcal{O}(N)$) asymptotic scaling were employed during the intermediate years \citep{1941ApJ....94..385H}.} Algorithmic advances and increasingly efficient hardware have enabled the simulation of globular clusters containing millions of stars \citep[e.g.,][]{2016MNRAS.458.1450W,2020MNRAS.497..536W}, in addition to long-duration integrations of our own solar system with gradually increasing accuracy \citep[e.g.,][]{1991AJ....102.1528W,2019MNRAS.490.5122R}.

A crucial ingredient in the simulation of dynamical systems is the method used to advance a calculation from one moment in time to the next. Predictor-corrector methods, first Adams-Bashforth-like multi-step methods \citep[e.g.,][]{1963MNRAS.126..223A,1973JCoPh..12..389A} and later methods that combined a Taylor series predictor with a 2-point Hermite corrector \citep[e.g.,][]{1991ApJ...369..200M,1999JCoAM.109..407S}, have been central to the study of collisional stellar (many-body) systems.\footnote{Rather than using Hermite or multi-step quadratures directly, many implementations use the resulting interpolant to approximate high-order derivatives of the acceleration which are then used to construct a high-order Taylor series. Such approaches can be advantageous for individual-timestep schemes.} High-order time-symmetric integrators \citep[e.g.,][]{1990AJ....100.1694Q}, as well as those based on Gaussian quadratures \citep[e.g.,][]{1985ASSL..115..185E,2015MNRAS.446.1424R}, have proven quite useful for conducting high-precision few-body simulations such as those of the Solar System. 

This work aims to enable more accurate and more efficient $N$-body simulations by introducing multi-step, multi-derivative schemes that generalize many of the aforementioned methods. Such methods have been studied for decades (see, for example, \citealt{Seal2014} for a brief review and an application to partial differential equations), but often from a mathematical rather than practical perspective. However, given the success of both multi-step and multi-derivative methods in astrophysics, multi-step multi-derivative methods are a natural extension. Perhaps most significantly, Section \ref{sec:variable} introduces a 3-point 6th-order extension of the classic 4th-order Hermite method that can enable more accurate and efficient simulations at negligible cost per timestep.

Section \ref{sec:math} introduces the mathematical framework of multi-point Hermite quadratures that underlies the schemes presented later, as well as some particular aspects of applying such methods to the gravitational $N$-body problem. Section \ref{sec:variable} introduces some 3-point variable-timestep predictor corrector methods of 6th, 9th, and 12th order (the lattermost is described in detail in Appendix \ref{app:3x4}).
Section \ref{sec:collocation} introduces a set of high-order time-symmetric quadratures that may be useful for long-term integrations of few-body systems, which are tested here in the form of implicit single-step collocation methods. Various formulae too cumbersome for the main text are deferred to the Appendices. Example symbolic computation scripts used to produce some the schemes introduced herein are available on GitHub.\footnote{\url{https://github.com/ajdittmann/HermiteScripts}}

\section{Mathematical Background}\label{sec:math}
\subsection{Hermite Interpolation and Integration}
Let us begin by considering a generic vector-valued ordinary differential equation 
\begin{equation}\label{eq:generic}
\frac{d\mathbf{y}}{dt}=f(\mathbf{y}(t),t).
\end{equation}
Given initial values $\mathbf{y}(t_0)$, the values at a later time $(t_1)$ are given by
\begin{equation}\label{eq:odeint}
\mathbf{y}(t_1) = \mathbf{y}(t_0) + \int_{t_0}^{t_1} f(\mathbf{y}(t),t) dt. 
\end{equation}
Simply approximating $f(\mathbf{y}(t),t)\approx f(\mathbf{y}(t_0),t_0)$ or $f(\mathbf{y}(t),t)\approx f(\mathbf{y}(t_1),t_1)$ leads to the first-order forward or backward Euler methods respectively. In general, we will construct higher-order integration methods by employing higher-order approximations of $f(\mathbf{y}(t),t)$ ($f(t)$ in the following).

The Hermite interpolant of $f(t)$, $\tilde{f}(t)$, is given by \citep[e.g.,][]{Hermite1877,16792fcc-ce52-33aa-ba02-aaef75cf8d37}
\begin{equation}\label{eq:herminterp}
\tilde{f}(t) = \sum_{j=0}^n \sum_{k=0}^r A_{jk}(t)f^{(k)}(t_j),
\end{equation}
where
\begin{align}
A_{jk}(t)=p_j(t)\frac{(t-t_j)^k}{k!}\sum_{m=0}^{r-k}\frac{1}{m!}(t-t_j)^m g^{(m)}_j(t_j),\\
p_j(t)=\frac{1}{(t-t_j)^{r+1}}\prod_{J=0}^n (t-t_J)^{r+1},\\
g_j(t) = p_j(t)^{-1}.
\end{align}
The notation $q^{(i)}$ denotes the $ith-$order total time derivative of some quantity $q$. In the above formulas, $n+1$ is the total number of points over which $f$ is interpolated, and $r$ is the order of the highest derivative of $f$ used in the interpolation.\footnote{For simplicity, I have assumed here (and throughout this work) that the same number of derivatives of $f$ are used at each point, but this need not be the case; $r$ would be replaced by $r_j$.} Hermite interpolation can be thought of an a generalization of Lagrange interpolation (and indeed, the above formulas reduce to the Lagrange interpolation formulas for $r=0$), but in addition to interpolating values of $f$ alone, some number of derivatives of $f$ are also interpolated across the interval. The degree of the resulting polynomial interpolant is $(r+1)(n+1)-1$.

The solution to Equation (\ref{eq:odeint}) can then be approximated as 
\begin{equation}\label{eq:intapprox}
\mathbf{y}(t_1)-\mathbf{y}(t_0)\approx \int_{t_0}^{t_1} \tilde{f}(t) dt,
\end{equation}
which yields an integration scheme with accuracy at least of order $(r+1)(n+1)$.

With $n=1$, defining $\Delta t = t_1-t_0$ and using $t_0$ and $t_1$ as interpolating points, Equation (\ref{eq:intapprox}) results in the trapezoidal rule for $r=0$, further compactifying notation such that $f_j\equiv f(t_j)$
\begin{equation}\label{eq:trap}
\mathbf{y}(t_1)-\mathbf{y}(t_0)\approx \frac{\Delta t}{2}\left(f_0 + f_1\right),
\end{equation}
the usual 4th-order Hermite method \citep[e.g.,][]{1991ApJ...369..200M,1992PASJ...44..141M} for $r=1$,
\begin{equation}\label{eq:herm4}
\begin{split}
\mathbf{y}(t_1)-\mathbf{y}(t_0)\approx\\ \frac{\Delta t}{2}\left(f_0 + f_1\right)+\frac{\Delta t^2}{12}\left(f^{(1)}_0 - f^{(1)}_1\right),
\end{split}
\end{equation}
the 6th-order method for $r=2$ \citep{https://doi.org/10.1002/sapm1953321171,2008NewA...13..498N}, 
\begin{equation}\label{eq:herm6}
\begin{split}
\mathbf{y}(t_1)-\mathbf{y}(t_0)\approx \frac{\Delta t}{2}\left(f_0 + f_1\right) \\ +\frac{\Delta t^2}{10}\left(f^{(1)}_0 - f^{(1)}_1\right) +\frac{\Delta t^3}{120}\left(f^{(2)}_0 + f^{(2)}_1\right),
\end{split}
\end{equation}
and the 8th-order method of \cite{2008NewA...13..498N} for $r=3$, 
\begin{equation}\label{eq:herm8}
\begin{split}
\mathbf{y}(t_1)-\mathbf{y}(t_0)\approx \\ \frac{\Delta t}{2}\left(f_0 + f_1\right) +\frac{3 \Delta t^2}{28}\left(f^{(1)}_0 - f^{(1)}_1\right) \\ + \frac{\Delta t^3}{84}\left(f^{(2)}_0 + f^{(2)}_1\right) + \frac{\Delta t^4}{1680}\left(f^{(3)}_0 - f^{(3)}_1\right).
\end{split}
\end{equation}
All of the above $n=1$ methods, and their extensions to higher $r$ \citep[e.g.,][]{8050caf2-dfa7-3e94-a6f3-4477cb4af95a,lanczos1956applied}, have the benefits of being time-symmetric and A-stable \citep[e.g.,][]{2021ApNM..160..205D}, the latter property making them suitable for solving stiff equations \citep{Dahlquist1963}.\footnote{Specifically, A-stable methods recover the correct asymptotic behavior as $t\rightarrow\infty$ when applied to the test problem $dy/dt=ky$ with ${\rm Re}\{k\}<0$. This property is also preserved when these schemes are applied in a predictor-corrector manner \citep[e.g.][]{2021ApNM..160..205D}.}

Setting $r=0$ and $n>1$ can lead to numerous classical multi-node collocation schemes and familiar multi-step integrators \citep[e.g.][]{ralston2001first,hairer2008solving}. As an example, let us take $n=3$ and $r=0$, sampling $f$ at four uniformly-spaced points. If we label these $t_j\in\{t_0,t_{1/3},t_{2/3},t_1\}$, and define $\Delta t \equiv t_1-t_0$, integrating from $t_0$ to $t_1$ yields the 4th-order Newton-Cotes rule (also known as Simpson's $3/8$ rule)
\begin{equation}\label{eq:nc4}
\mathbf{y}(t_1)-\mathbf{y}(t_0)\approx \frac{\Delta t}{8}\left(f_0+3f_{1/3}+3f_{2/3} + f_1\right).
\end{equation}

Relabeling the points as $t_j\in\{t_0,t_1,t_2,t_3\}$, redefining $\Delta t \equiv t_3-t_2,$ and integrating from $t_2$ to $t_3$ yields 
\begin{equation}\label{eq:am4}
\begin{split}
\mathbf{y}(t_3)-\mathbf{y}(t_2)\approx \frac{\Delta t}{24}\left(9f_3 + 19f_2 - 5f_1 + f_0 \right),
\end{split}
\end{equation}
the classic 4th-order Adams-Moulton formula \citep{bashforth1883attempt,moulton1926new}. Of course, it is not necessary to restrict ourselves to points within the domain of interpolation: for example, evaluating $f$ at the same points but integrating from $t_3$ to $t_4=t_3+\Delta t$ yields
\begin{equation}\label{eq:ab4}
\begin{split}
\mathbf{y}(t_4)-\mathbf{y}(t_3)\approx \frac{\Delta t}{24}\left(55f_3 - 59f_2 + 37f_1 - 9f_0 \right),
\end{split}
\end{equation}
the 4th-order Adams-Bashforth formula \citep{bashforth1883attempt}. Naturally, this extrapolation is only reliable in the immediate vicinity of where $f$ has been evaluated, leading to a much smaller region of stability for Equation (\ref{eq:ab4}) than for Equation (\ref{eq:am4}) \citep{Dahlquist1963}.

It is straightforward to extend the above integration methods to include higher derivatives of $f$, yielding higher-order schemes. For example, the 8th-order time-symmetric $r=1$ analog to Equation (\ref{eq:nc4}) is
\begin{equation}\label{eq:nch8}
\begin{split}
\mathbf{y}(t_1)-\mathbf{y}(t_0)\approx \\\frac{\Delta t}{244}\left(31f_0+81f_{1/3}+81f_{2/3} + 31f_1\right) \\
+ \frac{\Delta t^2}{3360}\left(19f_0^{(1)}-27f_{1/3}^{(1)}+27f_{2/3}^{(1)} -19f_1^{(1)}\right).
\end{split}
\end{equation}
Similarly, the $r=1$ analog of Equation (\ref{eq:am4}) is
\begin{equation}\label{eq:amh8}
\begin{split}
\mathbf{y}(t_3)-\mathbf{y}(t_2)\approx \\ \frac{\Delta t}{18144}\left(6893f_3 + 8451f_2 + 2403f_1 + 397f_0 \right)
+ \\\frac{\Delta t^2}{30240}\!\left(\!-1283f^{(1)}_3 \!\!+\! 7659f^{(1)}_2\!\! + \!2421f^{(1)}_1\!\! + \!163f^{(1)}_0 \!\right),
\end{split}
\end{equation}

and the $r=1$ analog of Equation (\ref{eq:ab4}) is
\begin{equation}\label{eq:abh8}
\begin{split}
\mathbf{y}(t_4)-\mathbf{y}(t_3)\approx \\ \frac{\Delta t}{18144}\left(-169325f_3 - 130275f_2 \right. \\\left.+ 256797f_1 + 60947f_0 \right)
+ \frac{\Delta t^2}{30240}\left(126467f^{(1)}_3 \right.\\\left.+ 490869f^{(1)}_2 + 315531f^{(1)}_1 + 25853f^{(1)}_0 \right).
\end{split}
\end{equation}
Although such methods are straightforward to derive and implement, I have not found a practical use for extrapolatory high-order multi-derivative analogs of Adams-Bashforth methods; in practice it seems that the limited stability of formulas such as Equation (\ref{eq:abh8}) is not conducive to their practical application, while purely interpolatory formulas such as Equations (\ref{eq:nch8}) and (\ref{eq:amh8}) typically produce both accurate and stable results.

\subsection{Gaussian-like Quadratures}
Oftentimes careful selection of node locations can improve accuracy. Gaussian quadratures select nodal locations such that polynomial integrands of as high an order as possible can be integrated exactly, sometimes subject to other constraints such as using the endpoints of the interval as quadrature nodes \citep[e.g.][]{gauß1814methodus,lobatto1852lessen}.\footnote{In the context of $N$-body simulations, these methods hold promise for few-body simulations or those where every particle shares the same time step. However, in simulations employing block or individual timesteps, these integrators are less useful because they rely on interpolants of relatively low degree compared to the overall order of the schemes, limiting their ability to predict the positions and velocities of passive particles.} For example, with $r=0$, the nodal locations of a Gaussian quadrature on the interval $[-1,1]$ can be found by solving the system
\begin{equation}
\int_{-1}^{1} t^k \prod_{J=0}^n (t-t_J) = 0, \hspace{1cm} k\in\{0,\ldots,n\}
\end{equation}
for each of $t_J$. Requiring the inclusion of the interval endpoints results in lower-order but often more convenient Gauss-Lobatto methods, the quadrature points for which can be found by solving the system
\begin{equation}\label{eq:glq}
\int_{-1}^{1} \!\!\!\!t^k (t-1)(t+1)\!\prod_{J=0}^{n-2} (t-t_J) = 0, \hspace{0.3cm} k\!\in\!\{0,\ldots,n-2\}.
\end{equation}
As an example, solving Equation (\ref{eq:glq}) with $n=2$ results in $t_J=0$, and upon transforming the interval $[-1,1]$ to $[t_0,t_1]$ leads to the Simpson's rule, 
\begin{equation}\label{eq:nc3}
\mathbf{y}(t_1)-\mathbf{y}(t_0)\approx \frac{\Delta t}{6}\left(f_0+4f_{1/2}+f_1\right),
\end{equation}
which achieves the same order of accuracy as Equation (\ref{eq:nc4}) with fewer function evaluations. Notably, the schemes resulting from Equation (\ref{eq:glq}) are time-symmetric; on the other hand, slightly higher orders of accuracy can be achieved with the same number of function evaluations by holding fewer nodes fixed, and high-order variants requiring only the nodality of the beginning of the interval \citep[Gauss-Radau schemes, after][]{radau1880etude} have proven popular for few-body integrations \citep[e.g.][]{1985ASSL..115..185E,2015MNRAS.446.1424R}.

Hermite interpolation is the basis of analogous schemes that utilize higher-order derivatives at all (or only a subset of) nodes \citep[e.g.][]{turan1950theory,stancu1963quadrature,50fe55be-56e2-3db2-a697-a00b366b969a}. In the case where $f$ and $r$ of its derivatives are employed at every node, quadrature points can be optimized by solving the system (again requiring the inclusion of the endpoints of the interval)
\begin{equation}\label{eq:glhq}
\int_{-1}^{1} \!\!\!\!t^k (t-1)^{r+1}\!(t+1)^{r+1}\!\!\prod_{J=0}^{n-2} (t-t_J)^{r+1} \!\!= \!0, \hspace{0.2cm} k\!\in\!\{0,\ldots,n-2\}.
\end{equation}
Unfortunately, this procedure is not as productive for odd values of $r$ \citep[e.g.][]{turan1950theory}, restricting its usefulness.\footnote{When $r$ is odd, the system of Equations (\ref{eq:glhq}) can only be satisfied by complex-valued solutions.}
In any case, once the quadrature points have been determined, Equation (\ref{eq:intapprox}) provides the weights for each point.

\subsection{Practical Approaches to Implicit Quadratures}
A potential limitation of many integration methods presented in the preceding sections is their implicit nature (for example, $\mathbf{y}(t_1)$ depending on $f(y(t_1))$  in Equation (\ref{eq:nc3})). One classic approach, employed by \cite{moulton1926new} for quadratures in the family of Equations (\ref{eq:am4}) and (\ref{eq:ab4}) is to use a lower-order explicit scheme to make an initial guess for the end-of-interval values needed to compute the higher-order formally-implicit quadrature. Alternatively, one may also attempt a truly implicit solution, iterating each quadrature until convergence.

\subsubsection{Predictor-Corrector Approaches}
The most common predictor-corrector method is perhaps the 2nd-order Runge-Kutta method \citep[e.g.,][]{runge1895numerische,heun1900neue}; that method employs a first-order explicit Euler step as a predictor followed by an application of the quadrature given by Equation (\ref{eq:trap}) approximating $f(y_1)$ as $f(y_0+\Delta t f(y_0))$. 

In the context of Hermite methods, which utilize higher-order derivatives of the fundamental differential equation, it is common to begin with a higher-order Taylor series generalization of the 1st-order explicit Euler predictor;\footnote{Taylor methods have also been applied directly to the $N$-body problem \citep[e.g.,][]{1974nsod.conf..451L,10.1093/mnras/stab1032}.} these Taylor series predictors are then corrected by the application of a Hermite corrector such as Equations (\ref{eq:herm4}-\ref{eq:herm8}) \citep[e.g.,][]{1992PASJ...44..141M,2008NewA...13..498N}. It is straightforward to construct a Taylor series predictor using the explicitly calculated values of $f$ and its derivative, although such a predictor will necessarily be two or more orders less accurate than the Hermite corrector. By evaluating the derivative of the underlying Hermite interpolant given by Equation (\ref{eq:herminterp}) at the end of each timestep, it is possible to construct higher-order Taylor series predictors for the next timestep.

\subsubsection{Fully Implicit Approaches}
Although predictor-corrector schemes are able to efficiently achieve high orders of accuracy, additional corrector applications are typically required to rigorously satisfy the corrector expressions (e.g. Equations \ref{eq:herm4}-\ref{eq:herm8}) and achieve the time-symmetry intrinsic to each quadrature when applicable. In some applications, time-symmetry can be achieved by applying the corrector multiple times, recomputing $f$ and its derivatives at each iteration \citep[e.g.,][]{1998MNRAS.297.1067K}. In some cases, particularly when solving stiff equations, this procedure may not converge if performed explicitly.

A more generally applicable method is to approach each timestep by solving the nonlinear equation (or system of equations) given by the underlying quadrature. The A-stable nature of 2-point Hermite schemes makes them particularly suitable for solving stiff equations, and they have been used to generate asymptotic-preserving implicit-explicit schemes of various orders \citep[e.g.,][]{SCHUTZ202184,2021ApNM..160..205D}. General multi-point quadratures can also be solved implicitly; as an example, Equation (\ref{eq:nch8}) can be applied by solving a nonlinear system of equations not just for $\mathbf{y}(t_1)$, but $\mathbf{y}(t_{1/3})$ and $\mathbf{y}(t_{2/3})$ as well, given the value $\mathbf{y}(t_{0})$. 

\subsection{Application to the Gravitational $N$-body Problem}
The $N$-body problem is not described only by Equation (\ref{eq:generic}), but also by $d^2\mathbf{x}/dt^2=\mathcal{G}(\mathbf{x}(t))$.
The trivial approach to this problem, which I will adopt in the remainder of this paper, is to simply evolve a system for both $\mathbf{y}$ and $d\mathbf{y}/dt$, evolving the positions and velocities of each body.\footnote{By using methods designed explicitly for second-order equations (particularly those of the form $d^2\mathbf{x}/dt^2=\mathcal{G}(\mathbf{x}(t))$), rather than approaching them as a set of first-order equations, higher-order schemes can be derived that potentially require fewer function evaluations \citep[e.g.,][]{stormer1921methode,1990AJ....100.1694Q}.} Although one could easily use higher-order quadratures for the positions, given the higher-order derivatives needed to update the velocities, the overall order of the scheme will often be limited by the accuracy of the velocity update.\footnote{Another approach is to use the extra freedom in each quadrature to better conserve other quantities. For example, when evolving periodic eccentric orbits using a time-symmetric integration scheme, errors can grow linearly in the argument of periapsis of the orbit despite the conservation of energy and angular momentum. One approach to numerically preserving this hidden symmetry \citep[see, for example,][]{Hamilton1844} is to apply different quadratures to the position and velocity of each body so as to cancel out the leading-order error in the evolution of the argument of periapsis \citep{2004PASJ...56..861K,2020MNRAS.496.1217D}.}

Explicitly, the governing equations, using units where the gravitational constant $G=1$, are
\begin{align}
\frac{d\mathbf{r}_i}{dt} = \mathbf{v}_i, ~~~~  \label{eq:dxdv}
\frac{d\mathbf{v}_i}{dt} = \mathbf{a}_i,  \\  \label{eq:accel}
\mathbf{a}_i = \sum_{j\neq i}\frac{m_j\mathbf{r}_{ij}}{\left(\mathbf{r}_{ij}\cdot\mathbf{r}_{ij}+\epsilon^2\right)^{3/2}},
\end{align}
where $\mathbf{r}_i$ and $\mathbf{v}_i$ are the position and velocity vectors of particle $i$ and $\mathbf{r}_{ij}\equiv\mathbf{r}_j-\mathbf{r}_i$. The gravitational softening parameter $\epsilon$ is an optional, typically unphysical, numerical crutch that prevents the denominator of Equation (\ref{eq:accel}) from approaching zero. Explicit expressions for some higher-order time derivatives of Equation (\ref{eq:accel}) are provided in Appendix \ref{sec:forces}. The errors of approximate solutions to the above equations are typically quantified using the total energy of the system, which would ideally be conserved and is given by 
\begin{equation}
E = \sum_i \frac{1}{2}m_i\mathbf{v}_i\cdot\mathbf{v}_i - \frac{1}{2}\sum_i\sum_{j\neq i}\frac{m_im_j}{\left(\mathbf{r}_{ij}\cdot\mathbf{r}_{ij}+\epsilon^2\right)^{1/2}}.
\end{equation}

As this system is governed by a Hamiltonian, certain classes of ``symplectic'' methods are often favored thanks to their preservation of phase-space structure \citep[see, e.g.,][for an overview]{1992AJ....104.1633S}. Unfortunately, the beneficial properties of these methods can easily be lost if they are applied using variable time steps; however, another important property of the above equations is their time-reversal symmetry, which is easier to preserve when using variable timesteps, though still nontrivial \citep[see, e.g.][]{1995ApJ...443L..93H}. Methods that respect this symmetry often conserve energy much better than generic integration schemes (see Appendix \ref{app:tsym} for a brief overview), and are discussed in Section \ref{sec:collocation}.

\subsubsection{The Choice of Time Step}
One of the challenges faced when solving the $N$-body problem, as with many differential equations, is choosing an appropriate time step. Although constant time steps may be suitable for near-circular orbits, they are not generally suitable for evolving highly-eccentric orbits; qualitatively, as the bodies move very quickly and forces are rather strong at pericenter small time steps are usually required, but at apocenter the system evolves comparatively slowly, such that using small time steps would waste considerable effort. If the timestep employed is too large to resolve the timescales physically relevant to the problem, the integration scheme will fail completely and produce order-unity errors, as for largest timesteps tested in Section \ref{sec:collocation}. As a basic example, when evolving a pair of objects with relative distance $r$ moving at relative velocity $\mathbf{v}$, a reasonable choice of time step might be $\Delta t = \eta \,r/v$, scaled by some dimensionless constant $\eta<1$. Most timestep criteria generalize this sort of reasoning to particle accelerations and their derivatives, avoiding division by zero. 

One of the most widely applied timestep criteria for collisional $N$-body simulations was empirically derived in \cite{AARSETH1985377}, defining the timescale
\begin{equation}\label{eq:aarseth}
\tau_{A} \equiv \eta\sqrt{\frac{|a||a^{(2)}|+|a^{(1)}|^2}{|a^{(1)}||a^{(3)}|+|a^{(2)}|^2}},
\end{equation}
in terms of the acceleration and three of its higher-order derivatives. Typically, this timescale is calculated for each particle and scaled by a dimensionless factor ($\eta$) to set an individual timestep for each particle. A straightforward approach when not simulating too many particles, which is often more efficient when performing parallelized computations, is to use a shared global timestep given by the minimum timescale out of all the particles; for larger-$N$ systems, grouping particles in constant-timestep blocks of various magnitudes is often advantageous \citep{1986LNP...267..156M}.

Equation (\ref{eq:aarseth}) can be easily generalized to include higher-order derivatives \citep[e.g.,][]{2008NewA...13..498N}. First defining
\begin{equation}
\mathcal{A}_k\equiv \sqrt{|a^{(k-1)}||a^{(k+1)}|+|a^{(k)}|^2},
\end{equation}
we can define a general timescale 
\begin{equation}\label{eq:genaarseth}
\tau_g \equiv \eta\left(\frac{\mathcal{A}_1}{\mathcal{A}_{p-2}}\right)^{1/(p-3)},
\end{equation}
which may offer benefits by incorporating the higher-order derivatives used by a given scheme. In Equation (\ref{eq:genaarseth}), $3<p<(r+1)(n+1)$ is an integer that can be at most one higher than the degree of the polynomial interpolant underlying a given method.

A lower-order approach, which is more robust to catastrophic roundoff errors,\footnote{Specifically, when timestep formulas rely on derivatives that are calculated using the underlying interpolant (as in Equations \ref{eq:c1}--\ref{eq:c6}), rather than being calculated directly, the estimated timesteps can be critically sensitive to roundoff errors. Out of the schemes tested in this work, this problem only occurs when applying Equation \ref{eq:genaarseth} to choose timesteps for the 3-point 12th-order scheme, as explored in Section \ref{app:3x4}.} was recently presented by \cite{2024OJAp....7E...1P}: 
\begin{equation}\label{eq:prs}
\tau_{P} = \eta\sqrt{\frac{2|a|^2}{|a||a^{(2)}|+|a^{(1)}|^2}}.
\end{equation}
Although many other timestep criteria are possible \citep[see, for example,][]{1991ApJ...369..200M,2008NewA...13..498N}, I have restricted myself to Equations (\ref{eq:aarseth}---\ref{eq:prs}) in the present study.

\section{Variable-Timestep Multistep Methods}\label{sec:variable}
Two-point 4th-order Hermite schemes are standard in modern collisional $N$-body calculations \citep[e.g,][]{2015MNRAS.450.4070W,2020MNRAS.497..536W}.\footnote{In this section, as in nearly all collisional simulations of star clusters, no attempt is made at time symmetry, and these schemes thus exhibit secular truncation-level error growth with time (see, for example, Appendix \ref{app:tsym}). Because the mutli-point methods introduced in this section can use larger timesteps to achieve a particular level of truncation error, they will accrue these secular errors more slowly than the standard 2-point schemes. } The 6th-order variation introduced by \cite{2008NewA...13..498N} has also been implemented in a number of codes \citep[e.g.,][]{2013JCoPh.236..580C,2015ComAC...2....8B}. In this section, I will show how these schemes can be extended to higher orders at the $\mathcal{O}(N)$ cost of more complicated predictor and corrector procedures, \textit{but the same number of force evaluations.} These gains are achieved by utilizing force evaluations from previous timesteps.

\begin{figure}[h]
\begin{centering}
\begin{tikzpicture}
\hspace{1.2cm}
\draw[very thick,-{latex}] (0,0) -- (6,0) node[below]{$t$};
\foreach \x/\l in {1/_{-1},2.5/{_0},5/_1}
  \draw (\x,3pt) -- (\x,-3pt) node[below]{$t\l$};
\draw [decorate,decoration={brace,amplitude=5pt,raise=1.5ex}]
  (1,0) -- (2.5,0) node[midway,yshift=1.75em]{$\Delta t_0$};
\draw [decorate,decoration={brace,amplitude=5pt,raise=1.5ex}]
  (2.5,0) -- (5,0) node[midway,yshift=1.75em]{$\Delta t_1$};
\end{tikzpicture}
\end{centering}
\caption{A schematic diagram illustrating the points in time used by the 3-point variable-timestep Hermite schemes and the notation used in this section. To simplify notation, I define $\zeta\equiv\Delta t_0/\Delta t_1$.}\label{fig:schematic1}
\end{figure}

In general, the goal will be to find a high-order approximate solution to the particle positions and velocities at time $t_1$ given evaluations of the relevant forces and their derivatives at $t_0$ and $t_{-1}$, defining $\Delta t_1\equiv t_1-t_0$ and $\Delta t_0\equiv t_0-t_{-1}$ (see Figure \ref{fig:schematic1}). Defining $\zeta\equiv \Delta t_0/\Delta t_1$, we can write a 3-point 6th-order corrector for the velocity vector of each particle as 
\begin{equation}\label{eq:var32}
\mathbf{v}_1 - \mathbf{v}_0 = \Delta t_1 \sum_{i=-1}^1c_{i,0}\mathbf{a}_i + \Delta t_1^2 \sum_{i=-1}^1c_{i,1}\mathbf{a}^{(1)}_i,
\end{equation}
in terms of its acceleration $\mathbf{a}$ and the first time-derivative of the acceleration $\mathbf{a}^{(1)}$. The values of $c_{ij}$ in Equation (\ref{eq:var32}) are given by 
\begin{equation}
\begin{split}
c_{-1,0} &= \frac{1}{30}\frac{5\zeta^2 + 5\zeta +1}{\zeta^3(\zeta^3 + 3\zeta^2 + 3\zeta +1)}\\
c_{0,0} &= \frac{1}{30}\frac{15\zeta^3 + 4\zeta^2 - 2\zeta - 1}{\zeta^3}\\
c_{1,0} &= \frac{1}{30}\frac{15\zeta^3 + 41\zeta^2 + 35\zeta +10}{\zeta^3 + 3\zeta^2 + 3\zeta +1}\\
c_{-1,1} &= \frac{1}{60}\frac{2\zeta+1}{\zeta^2(\zeta^2 + 2\zeta +1)}\\
c_{0,1} &= \frac{1}{60}\frac{5\zeta^2+4\zeta+1}{\zeta^2}\\
c_{1,1} &= \frac{1}{60}\frac{-5\zeta^2-6\zeta-2}{\zeta^2+2\zeta+1}.
\end{split}
\end{equation}
These expression follow from Equations (\ref{eq:herminterp}---\ref{eq:intapprox}), setting $n=2$ and $k=1$. In the notation discussed above, a quadrature is constructed using nodes at $t_{-1}$, $t_{0}$, and $t_1$, which is then integrated from $t_0$ to $t_1$.

In order to employ a high-order Taylor series predictor alongside the above 3-point Hermite corrector, it is useful to have estimates for higher-order derivatives of $\mathbf{a}$ than those used directly in Equation (\ref{eq:var32}). Expressions for $\mathbf{a}^{(2)}_2$---$\mathbf{a}^{(5)}_2$ computed in this manner can be found in Appendix \ref{app:3x2derivs}. The corrector for the position of each particle $\mathbf{x}$ can be found by replacing $\mathbf{a}$ and its derivatives with $\mathbf{v}$ and its derivatives.
Equipped with these approximations, predicted values of velocities at $t_1$ ($\tilde{\mathbf{v}}_1$) can be determined using a simple Taylor series
\begin{equation}
\tilde{\mathbf{v}}_1 = \mathbf{v}_0 + \Delta t \mathbf{a}_0 + \sum_{k=2}^5\frac{\Delta t_1^k}{k!}a^{(k-1)}_0,
\end{equation}
after which $\tilde{\mathbf{v}}_1$ and the analogous approximations of $\tilde{\mathbf{x}}_1$ can be used to approximate $\mathbf{a}_1$ and $\mathbf{a}^{(1)}_1$ in Equation (\ref{eq:var32}). 

A 3-point 9th-order generalization of Equation (\ref{eq:herm6}) and Equation (\ref{eq:var32}) can be expressed as
\begin{equation}\label{eq:var33}
\begin{split}
\mathbf{v}_1 - \mathbf{v}_0 = \Delta t_1 \sum_{i=-1}^1d_{i,0}\mathbf{a}_i + \Delta t_1^2 \sum_{i=-1}^1d_{i,1}\mathbf{a}^{(1)}_i \\+\Delta t_1^3 \sum_{i=-1}^1d_{i,2}\mathbf{a}^{(2)}_i,
\end{split}
\end{equation}
where

\begin{align}
\begin{split}
d_{-1,0} &= \frac{1}{420}\frac{-84\zeta^4 - 168\zeta^3 - 124\zeta^2 - 40\zeta - 5}{\zeta^5(\zeta^5 + 5\zeta^4 + 10\zeta^3 + 10\zeta^2 + 5\zeta +1)}\\
d_{0,0} &= \frac{1}{420}\frac{210\zeta^5 + 54\zeta^4 - 27\zeta^3 - \zeta^2 + 15\zeta +5}{\zeta^5}\\
d_{1,0} &= \frac{210\zeta^5 \!\!+\! 996\zeta^4\!\! +\! 1857\zeta^3 \!\!+\! 1696\zeta^2\! \!+\! 770\zeta \!+\! 140}{420(\zeta^5 + 5\zeta^4 + 10\zeta^3 + 10\zeta^2 + 5\zeta +1)}\\
d_{-1,1} &= \frac{1}{840}\frac{-42\zeta^3 - 63\zeta^2 - 31\zeta - 5}{\zeta^4(\zeta^4 + 4\zeta^3 + 6\zeta^2 + 4\zeta +1)}\\
d_{0,1} &= \frac{1}{840}\frac{84\zeta^4 + 54\zeta^3 - 9\zeta^2 - 19\zeta - 5}{\zeta^4}\\
d_{1,1} &= \frac{1}{840}\frac{-84\zeta^4 - 282\zeta^3 - 333\zeta^2 - 175\zeta - 35}{\zeta^4 + 4\zeta^3 + 6\zeta^2 + 4\zeta +1}\\
d_{-1,2} &= \frac{1}{5040}\frac{-18\zeta^2 - 18\zeta - 5}{\zeta^3(\zeta^3 + 3\zeta^2 + 3\zeta +1)}\\
d_{0,2} &= \frac{1}{5040}\frac{42\zeta^3 + 54\zeta^2 + 27\zeta +5}{\zeta^3}\\ 
d_{1,2} &= \frac{1}{5040}\frac{42\zeta^3 + 72\zeta^2 + 45\zeta +10}{\zeta^3 + 3\zeta^2 + 3\zeta +1}.
\end{split}
\end{align}
Expressions for $\mathbf{a}^{(3)}_2$---$\mathbf{a}^{(8)}_2$ computed by evaluating derivatives of this interpolant can be found in Appendix \ref{app:3x3derivs}. 

Because the 8th-order Hermite scheme of \cite{2008NewA...13..498N} has not proven as popular as the 4th-order and 6th-order 2-point schemes and because the corresponding expressions are rather long, I have relegated the 3-point 12th-order Hermite corrector to Appendix \ref{app:3x4}, although it is marginally more efficient at reaching roundoff-limited accuracy than the 9th-order scheme tested below. The corresponding expressions for $\mathbf{a}^{(4)}$---$\mathbf{a}^{(11)}$ can be found in in Appendix \ref{app:3x4derivs}.

\subsection{Convergence Tests}
A number of tests of the preceding methods follow, employing each method in conjunction with one of the time stepping formulas introduced earlier: Equation (\ref{eq:aarseth}) (``Aarseth''); Equation (\ref{eq:genaarseth}), using the highest-order derivatives of a given scheme (``generalized Aarseth,'' setting $p$ in Equation \ref{eq:genaarseth} equal to the order of each scheme); and Equation (\ref{eq:prs}) (``PRS''), which utilizes lower-order derivatives. After calculating $\Delta t_i$ for each particle using the aforementioned equations, the minimum timestep was determined and used to advance all of the particles in unison.

The error of each simulation was measured by calculating the energy of the system at the beginning of each simulation ($E_0\equiv E(t=0)$) at the end of every time step measuring the deviation between the current and initial energy ($\Delta E\equiv E-E_0$), and then identifying the maximum error over the course of each simulation. Convergence is reported with respect to two quantities: the dimensionless prefactor $\eta$ which scales the timesteps prescribed by Equations (\ref{eq:prs}---\ref{eq:aarseth}), and the number of total force evaluations $N_{\rm eval}$, which is proportional to the inverse of the average time step and represents the main cost of larger-$N$ simulations. 

For these 3-point quadratures, there is also the question of initialization. A few approaches are possible, such as guessing values at $t_{-1}$ and $t_{1}$ based on those at $t_0$ and correcting them implicitly, which I adopt in Section \ref{sec:2bodyvar}. 
Another approach is to use some other single-step integration scheme with a sufficiently small step size as to result in subdominant errors. In section \ref{sec:plummer} I employ a 4th-order Taylor method and subdivide the initial two intervals (determined using the appropriate time-step criterion for the Hermite scheme) into ten parts, reaching machine precision.

\subsubsection{A 2-Body System}\label{sec:2bodyvar}
\begin{figure}
\includegraphics[width=\linewidth]{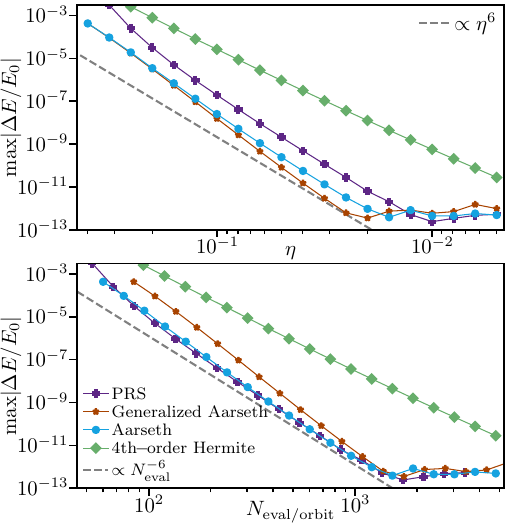}
\caption{The convergence of the 3-point 6th-order variable-timestep scheme with respect to the dimensionless timestep control $\eta$ and the number of force evaluations during a 100-orbit eccentric 2-body test problem. Each scheme converges at the expected rate, or perhaps slightly faster when taking larger timesteps. The different timestep criteria result in minor differences in performance, though the generalized Aarseth criterion is slightly less efficient at lower accuracies.}\label{fig:3x2}
\end{figure}
This subsection presents a suite of proof-of-concept simulations of a 2-body system with mass ratio $q=10^{-4}$ and eccentricity $e=0.9$. The system was integrated for 100 orbits. For simplicity, each of these simulations used the highest-possible-order Taylor series to predict the position and velocity of each particle before applying a Hermite corrector. The accelerations and higher-order derivatives thereof required by each corrector were calculated directly, and only the positions and velocities of each particle were predicted.\footnote{Because $\mathbf{a}^{(2)}$ depends on $\mathbf{a}$ and $\mathbf{a}^{(3)}$ depends on both $\mathbf{a}$ and $\mathbf{a}^{(1)}$, large-scale performance-conscious simulations would likely predict $\mathbf{a}$ and $\mathbf{a}^{(1)}$ to better parallelize the calculation, potentially at the expense of some accuracy. For this reason, the results presented here for the 3-point 9th-order scheme, and in Appendix \ref{app:3x4conv} for the 3-point 12th-order scheme, may be slightly more accurate than a more performant implementation, although the overall order of the scheme would be unaffected.} For this problem, the gravitational softening length was set to $\epsilon=0$. 

\begin{figure}
\includegraphics[width=\linewidth]{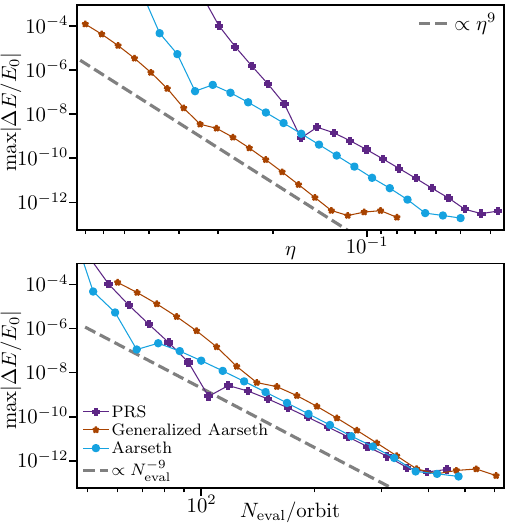}
\caption{The convergence of the 3-point 9th-order variable-timestep scheme with respect to the dimensionless timestep control $\eta$ and the number of force evaluations in a 100-orbit eccentric 2-body test problem. In this case the PRS and Aarseth timestep criteria typically perform better than the generalized Aarseth criterion, resulting in fewer force evaluations to achieve a particular degree of accuracy.}\label{fig:3x3}
\end{figure}

Results for the 3-point 6th-order scheme are presented in Figure \ref{fig:3x2}. The scheme converges at slightly faster than 6th order with respect to $\eta$ at larger time steps. Although the PRS criterion requires smaller values of $\eta$ to reach a given level of error, it performs similarly to the classical Aarseth criterion. For this problem, the 6th-order scheme requires roughly $\sim7\times10^2$ force evaluations per orbit to achieve roundoff-dominated errors. Results computed using the 2-point 4th-order Hermite scheme, using the Aarseth timestep criterion, are shown for comparison; for a given $\eta$ or number of force evaluations, the 3-point scheme is unequivocally superior.

Figure \ref{fig:3x3} presents analogous results for the 3-point 9th-order scheme, which generally converges as expected. In this case the generalized Aarseth criterion is less efficient than the other timestep criteria. The PRS criterion requires smaller values of $\eta$ to resolve the orbit to an acceptable degree, and is slightly more efficient for achieving errors below $\sim10^{-8}$; the Aarseth criterion is more efficient for larger error tolerances. For this problem, the 9th-order scheme required roughly $\sim4\times10^{2}$ force evaluations per orbit to achieve roundoff-dominated errors. Thus, to reach machine-limited accuracy even without parallelization, the 9th-order scheme will typically be more efficient than the 6th- or lower-order schemes, despite the additional floating point operations incurred by calculating higher-order derivatives of the acceleration \citep[see, for example, Table 1 of ][]{2008NewA...13..498N}.

\subsubsection{A Plummer Model}\label{sec:plummer}
Having confirmed the convergence of the variable-timestep schemes when applied to a toy problem, we turn now to a test problem closer to a large-scale cluster simulation. Below, I present the results for a short-timescale (to $t=10$ in standard units \citep{1971Ap&SS..14..151H} where $G$ and the system mass are unity, and the system energy is $-1/4$) direct $N$-body simulation of a 1024-object Plummer sphere, largely following \citep{2008NewA...13..498N}. In this test the gravitational softening length was set to $\epsilon=4/N$, comparable to the average inter-particle spacing. Each simulation used identical initial conditions, {and the same predictor-corrector approach employed in Section \ref{sec:2bodyvar}}. 

\begin{figure}
\includegraphics[width=\linewidth]{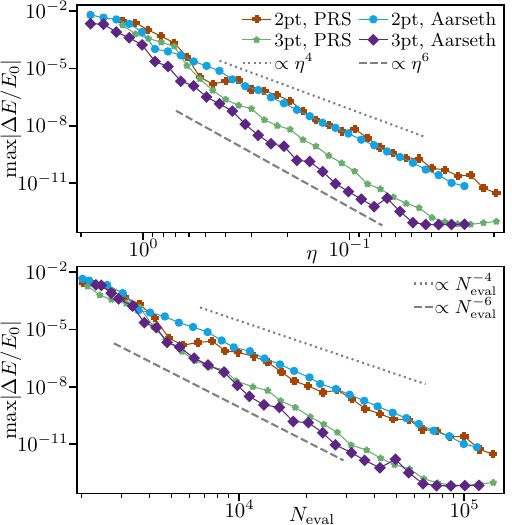}
\caption{The energy errors incurred over the course of short-term (until $t=10$ in H\'{e}non units) simulations of a 1024-object Plummer sphere using 2-point and 3-point Hermite integrators. The higher-order scheme is unequivocally more efficient, although the improvements it yields are more substantial at error tolerances below $\sim10^{-6}$ or $\sim10^{-4}$ depending on the timestep criterion employed.}\label{fig:cluster}
\end{figure}

Because of the widespread popularity of the 2-point 4th-order Hermite method, I have focused here on a comparison between that scheme and the 3-point 6th-order Hermite method, using the PRS and Aarseth timestep criteria. Because these two integrators use the same evaluations of the force and its derivatives, their cost only differs by the $\propto\mathcal{O}(N)$ costs such as correctors, the slightly higher memory footprint of the 3-point scheme, and the number of force evaluations required to achieve a given accuracy.  

The results of these tests are plotted in Figure \ref{fig:cluster}. 
The Aarseth timestep was typically slightly more efficient for this test problem, although the PRS criterion often produced comparable results. The 3-point method was unequivocally more efficient than the 2-point method regardless of the employed timestep criterion, in terms of both $\eta$ and the requisite number of force evaluations. The benefits of the 3-point scheme were marginal for simulations with fractional errors above $\sim 10^{-4}$, but the sixth-order scheme required a factor of $\sim 3$ fewer force evaluations to achieve an error of $\sim10^{-8}$. Thus, these sixth-order schemes will also enable simulations to use larger timesteps than their 4th-order counterparts, enabling faster simulations. 

\section{Single-Step Collocation Methods}\label{sec:collocation}
The mathematical formalism discussed in Section \ref{sec:math} can also be used to generate single-step methods that utilize multiple nodes throughout the timestep. A common example of such a method is the popular 4th-order Runge-Kutta integrator \citep{Kutta}, which uses explicit estimates to approximate Equation (\ref{eq:nc4}). In the following, to avoid developing and tuning a large number of Runge-Kutta schemes, I have employed each quadrature implicitly. 
I have restricted my focus here to time-symmetric integrators that include both the beginning and end of each time step as nodes, although following Section \ref{sec:math} it is trivial to loose these constraints. Some of the benefits of time-symmetric schemes are illustrated and discussed further in Appendix \ref{app:tsym}. In the following subsection I have used constant timesteps, although as single-step methods these schemes can trivially accommodate variable time stepping.

\begin{figure}[h!]
\begin{centering}
\begin{tikzpicture}
\hspace{1.2cm}
\draw[very thick,-{latex}] (0,0) -- (6,0) node[below]{$t$};
\foreach \x/\l in {1/_0,3/{_{1/2}},5/_1}
  \draw (\x,3pt) -- (\x,-3pt) node[below]{$t\l$};
\draw [decorate,decoration={brace,amplitude=5pt,raise=5.5ex}]
  (1,0) -- (3,0) node[midway,yshift=3.25em]{$\Delta t/2$};
\draw [decorate,decoration={brace,amplitude=5pt,raise=1.5ex}]
  (1,0) -- (5,0) node[midway,yshift=1.75em]{$\Delta t$};
\end{tikzpicture}
\end{centering}
\caption{A schematic diagram illustrating the points in time used by the $n=2$, $r=1$ time-symmetric Hermite schemes and the notation used in this section. Equation (\ref{eq:3x2mid}) integrates from $t_0$ to $t_{1/2}$, while Equation (\ref{eq:sym3x2}) integrates from $t_0$ to $t_1$, though both use the same underlying interpolant.}\label{fig:schematic2}
\end{figure}

\begin{figure*}
\includegraphics[width=\linewidth]{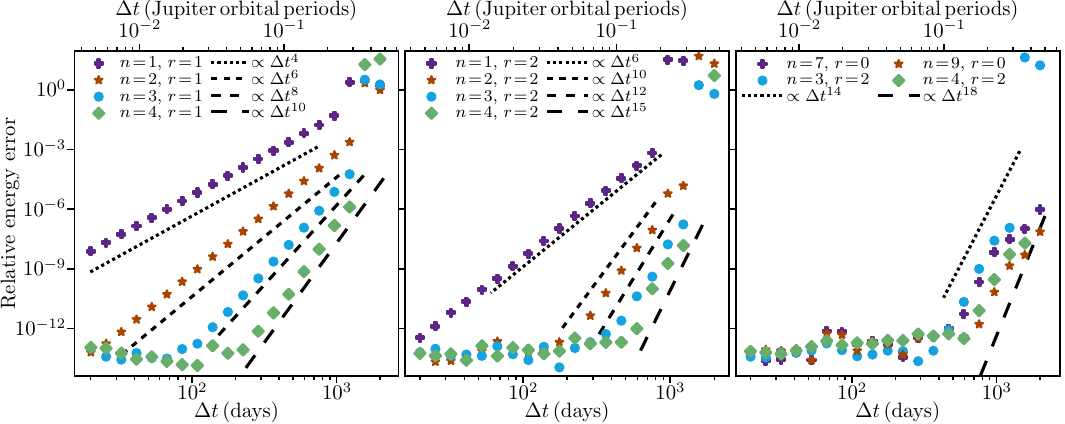}
\caption{The convergence of a variety of time-symmetric methods when applied to $10^4$-year integrations of the outer solar system. The right panel plots the results of $(n+1)$-point 2-derivative schemes; the purple crosses in that panel are analogous to the time-symmetric adaption of the 2-point Hermite schemes \citep[e.g.][]{1998MNRAS.297.1067K}. The central panel plots similar convergence tests for the $(n+1)$-point 3-derivative schemes. The rightmost panel illustrates the convergence of Gaussian-like schemes, comparing 8- and 10-point Gauss-Lobatto quadratures (plotted using purple crosses and orange stars) to the analogous 4- and 5-point $r=2$ Gaussian-like quadratures (plotted using blue circles and green diamonds). }\label{fig:alltimesym}
\end{figure*}

Taking the $n=2$, $r=1$ quadrature as an example, Figure \ref{fig:schematic2} illustrates how the time step is discretized. 
In order to solve implicitly for $\mathbf{y}(t_1)$ given only its derivatives at $t_0$, one must also solve for $\mathbf{y}(t_{1/2})$. The quadratures in this case, using the velocity as an example, are given by
\begin{align} \label{eq:sym3x2}
\mathbf{v}_{1/2}=\mathbf{v}_0 + \frac{\Delta t}{480}\left(101\mathbf{a}_0 + 128\mathbf{a}_{1/2} + 11\mathbf{a}_1\right)\\ \nonumber + \frac{\Delta t^2}{960}\left(13\mathbf{a}_0^{(1)} - 40\mathbf{a}_{1/2}^{(1)} - 3\mathbf{a}_1^{(1)}\right) \label{eq:3x2mid}\\ 
\mathbf{v}_{1} = \mathbf{v}_0 + \frac{\Delta t}{30}\left(7\mathbf{a}_0 + 16\mathbf{a}_{1/2} + 7\mathbf{a}_1 \right) \\ \nonumber + \frac{\Delta t^2}{60}\left(\mathbf{a}_0^{(1)}-\mathbf{a}_1^{(1)}\right).
\end{align}
Additional correctors for schemes using derivatives as high as $\mathbf{a}^{(2)}$, using between three and five points on equally subdivided intervals are provided in Appendix \ref{app:timesym}. In the following I will also compare two Gauss-Lobatto methods to a pair of $r=2$ Gaussian-like Hermite methods of similar order, also fixing the beginning and end of each interval as nodes; the corresponding formulae are provided in Appendix \ref{app:gaussian}. As a word of caution, unlike the schemes discussed in the previous section, some of these schemes (those with $n\geq2$ and equally-spaced nodes) would be unstable if applied as multi-step methods rather than single-step methods \citep[e.g.][]{10.1093/imanum/18.1.57,1999AJ....118.1888E}. As a reminder from Section \ref{sec:math}, in this notation $n+1$ points are used in the quadrature, and $r$ is the highest-order derivative employed in each quadrature (thus $r=0$ signifies using the acceleration when updating the velocity).

\subsection{Convergence Tests}
I have conducted a number of $10^4$-year integrations of the outer solar system to test the convergence of these schemes. At each time step, the system was advanced implicitly: each step began by using all of the explicitly-calculated derivatives available at $t_0$ to predict particle positions and velocities at $t_{i/(n+1)},\,i\in\{1,...,n+1\}$ using a Taylor series; the corrector was then iterated to convergence (a tolerance of $10^{-14}$) using the MINPACK \texttt{hybrj} algorithm \citep{1978LNM...630..105M,2020SciPy-NMeth}, reevaluating forces and the derivatives thereof as necessary; the corrector was implemented using \texttt{jax} \citep{jax2018github}, and the Jacobian of each corrector was calculated using automatic differentiation.

For the most part, the convergence tests of these schemes, presented in Figure \ref{fig:alltimesym}, hold few surprises; the equally-subdivided schemes converge at their theoretical orders of convergence, save for when the schemes become unstable at larger time steps (often between half and a fifth of Jupiter's orbital period). Naturally, schemes that subdivide each time step into larger numbers of intervals can stably accommodate larger maximum time steps. Note that in the middle panel of Figure \ref{fig:alltimesym}, the $n=2, r=2$ scheme converges at 10th order rather than 9th because in that case the equally-spaced quadrature is also a Gaussian-like quadrature.

The integration schemes based on multiderivative Gaussian quadratures converge at roughly the same order as the Gauss-Lobatto schemes, but produce results with very slightly larger errors at a given time step. The 14th-order Gauss-Lobatto scheme used here is roughly analogous to the 15th-order Gauss-Radau scheme used by \cite{2015MNRAS.446.1424R}, although that scheme sacrifices time symmetry for an additional order of accuracy. At constant order the multiderivative schemes gain fewer orders of accuracy through optimized node placement and are underlain by a higher-order interpolant, which could enable more accurate dense output or better predictor steps (and thus faster convergence in some applications). 

These time-symmetric multiderivative schemes may improve the efficiency of long-duration simulations of few-body systems. They have the potential to be particularly beneficial when parallelized, since the computation of higher-order derivatives can increase the number of floating point operations conducted per memory access.\footnote{One popular parallelization strategy for simulating large systems of particles is to distribute the calculation of force (and its derivatives) across processors. In smaller-N simulations, particularly those solving the underlying quadrature implicitly, it may be advantageous to distribute the force calculation at each node between cores.} However, the additional expense incurred when calculating $\mathbf{a}^{(j+1)}$ compared to $\mathbf{a}^{(j)}$ makes the preferability of these schemes nontrivial when applied in serial. For the purely gravitational $N$-body problem, the number of floating point operations required to calculate $\mathbf{a}$---$\mathbf{a}^{(2)}$ can be roughly $\approx2.6$ times the operation count required to calculate $\mathbf{a}$ alone \citep[e.g.,][]{2008NewA...13..498N}; thus, employing higher-order derivatives has the potential to increase the order of the scheme more efficiently than adding more quadrature nodes. However, considering the additional orders of accuracy to be gained by optimal node placement, multiderivative Gaussian-like schemes tend to be disfavored in terms of raw floating-point operations, but offer potential benefits in terms of parallelization and the higher-order predictors they enable.

\section{Summary}

This work has presented a family of multi-step multi-derivative $N$-body integrators. Conceptually, these methods can be thought of as combining aspects of both the Adams-Bashforth-Moulton and Hermite families of methods, both of which are special cases of the methods discussed here. Section \ref{sec:math} introduced the mathematical basis for these schemes, which was used to derive the schemes tested subsequently and can easily be extended to cases left unexplored in this work.

Perhaps the most promising immediate application of these methods is to large-$N$ collisional $N$-body simulations, where the 3-point 2- or 3-derivative Hermite methods discussed in Section \ref{sec:variable} can make simulations of star clusters more accurate at negligible ($\mathcal{O}(N)$) additional cost. In tests of both eccentric Keplerian orbits and Plummer spheres, the 3-point schemes outperform the 2-point schemes; the higher-order schemes achieve lower errors at a given number of force evaluations in all cases, but their benefits are most pronounced when attempting to achieve relative errors of $\sim10^{-4}$ or lower. 

Section \ref{sec:collocation} introduced a number of time-symmetric schemes suitable for long-term integrations of few-body systems. These schemes have the potential to facilitate somewhat more efficient integrations than acceleration-only schemes of the same orders thanks to their amenability to parallelization.


\section*{Acknowledgments}
I am very grateful for the insightful comments of Jun Makino on a preliminary version of this manuscript. I am also thankful for useful discussions with Hanno Rein and Ryan Westernacher-Schneider.
Support for this work was provided by NASA through the NASA Hubble Fellowship grant
\#HST-HF2-51553.001 awarded by the Space Telescope Science Institute, which is operated by the Association of Universities for Research in Astronomy, Inc., for NASA, under contract NAS5-26555. The early stages of this work were supported by NASA ADAP grants 80NSSC21K0649 and 80NSSC20K0288. 

\pagebreak

\appendix
\onecolumn
\section{Explicit Expressions for Time Derivatives of Acceleration}\label{sec:forces}
Using the compact notation of \cite{2008NewA...13..498N}, the acceleration of particle $i$ due to its gravitational interaction with particle $j$, $\mathbf{A}_{ij}$, and its first three time derivatives ($\mathbf{J}_{ij}$, $\mathbf{S}_{ij}$, $\mathbf{C}_{ij}$), are
\begin{align}
\mathbf{A}_{ij}&=m_j\frac{\mathbf{r}_{ij}}{\left(\mathbf{r}_{ij}\cdot\mathbf{r}_{ij}+\epsilon^2\right)^{3/2}}\\
\mathbf{J}_{ij}&=m_j\frac{\mathbf{v}_{ij}}{\left(\mathbf{r}_{ij}\cdot\mathbf{r}_{ij}+\epsilon^2\right)^{3/2}} - 3\alpha \mathbf{A}_{ij};  &\alpha=&\frac{\mathbf{r}_{ij}\cdot\mathbf{v}_{ij}}{\left(\mathbf{r}_{ij}\cdot\mathbf{r}_{ij}+\epsilon^2\right)}\\
\mathbf{S}_{ij}&=m_j\frac{\mathbf{a}_{ij}}{\left(\mathbf{r}_{ij}\cdot\mathbf{r}_{ij}+\epsilon^2\right)^{3/2}} - 6\alpha\mathbf{J}_{ij}-3\beta\mathbf{A}_{ij}; &\beta=&\frac{\mathbf{v}_{ij}\cdot\mathbf{v}_{ij}+\mathbf{r}_{ij}\cdot\mathbf{a}_{ij}}{\left(\mathbf{r}_{ij}\cdot\mathbf{r}_{ij}+\epsilon^2\right)}+\alpha^2\\
\mathbf{C}_{ij}&=m_j\frac{\mathbf{j}_{ij}}{\left(\mathbf{r}_{ij}\cdot\mathbf{r}_{ij}+\epsilon^2\right)^{3/2}} - 9\alpha\mathbf{S}_{ij}-9\beta\mathbf{J}_{ij}-3\gamma\mathbf{A}_{ij}; &\gamma=&\frac{3\mathbf{v}_{ij}\cdot\mathbf{a}_{ij}+\mathbf{r}_{ij}\cdot\mathbf{j}_{ij}}{\left(\mathbf{r}_{ij}\cdot\mathbf{r}_{ij}+\epsilon^2\right)}+\alpha(3\beta-4\alpha^2),
\end{align}
where $\mathbf{v}_{ij}=\mathbf{v}_j-\mathbf{v}_i$, $\mathbf{a}_{ij}=\mathbf{a}_j-\mathbf{a}_i$, and $\mathbf{j}_{ij}=\mathbf{a}^{(1)}_j-\mathbf{a}^{(1)}_i$. Using the above expressions, the acceleration of each particle and its derivatives are simply
\begin{equation}
\mathbf{a}_i=\sum_j \mathbf{A}_{ij};~~~~
\mathbf{a}^{(1)}_i=\sum_j \mathbf{J}_{ij};~~~~
\mathbf{a}^{(2)}_i=\sum_j \mathbf{S}_{ij};~~~~
\mathbf{a}^{(3)}_i=\sum_j \mathbf{C}_{ij}.
\end{equation}
\section{Time-Symmetric Methods}\label{app:tsym}
Newtonian gravitational dynamics conserves energy and exhibits time-reversal symmetry. Numerical schemes for the N-body problem that discretely preserve this symmetry often exhibit lower long-term energy errors than comparable asymmetric schemes
\citep[e.g,][]{{1990AJ....100.1694Q},{1995ApJ...443L..93H},{Evans_1999},{2006NewA...12..124M}}.
When integrating periodic orbits such as Keplerian ellipses, the truncation errors of time-symmetric schemes lead to energy errors bounded in time; integrations of periodically dominant but asymmetrically perturbed systems typically accrue secular truncation errors related to the strength of their perturbations \citep[see, e.g.,][]{Hairer_2006,10.1093/mnras/sty184}. Thus, schemes along the lines of those explored in Section \ref{sec:variable} accrue secular truncation-level errors proportional to the simulation time, while schemes such as those explored in Section \ref{sec:collocation} can result in bounded truncation errors. In both cases integrations accumulate rounding errors, which are typically subdominant in the former case but can be the limiting factor in the latter and typically grow proportionally $\propto t$ or $\sqrt{t}$ depending on biases or lack thereof in floating point arithmetic \citep{1937AJ.....46..149B}.\footnote{For example, Figure 6 of \citep{2015MNRAS.446.1424R} shows that many implementations of symplectic integrators accrue energy error proportional to time, whereas their integrator accrues errors at much closer to a random walk (see, however, \citealt{2021MNRAS.502..556H}).}

For an integration scheme to benefit from a time-symmetric underlying quadrature, its timestep should be chosen in a time symmetric manner. As an example, let us consider a quadrature such as Equation (\ref{eq:herm6}): clearly the equation is unchanged upon exchanging the indices $1$ and $0$ and changing the sign of $\Delta t$. Although this is trivial with constant-magnitude timesteps, it can be plunged into jeopardy if its variable timesteps are chosen in a time-asymmetric manner. While time-symmetric variable-timestep schemes are relatively simple to construct, their cost tends to make them impractical \citep[see, e.g.,][]{1995ApJ...443L..93H,2006NewA...12..124M}.

\begin{figure}
\includegraphics[width=\linewidth]{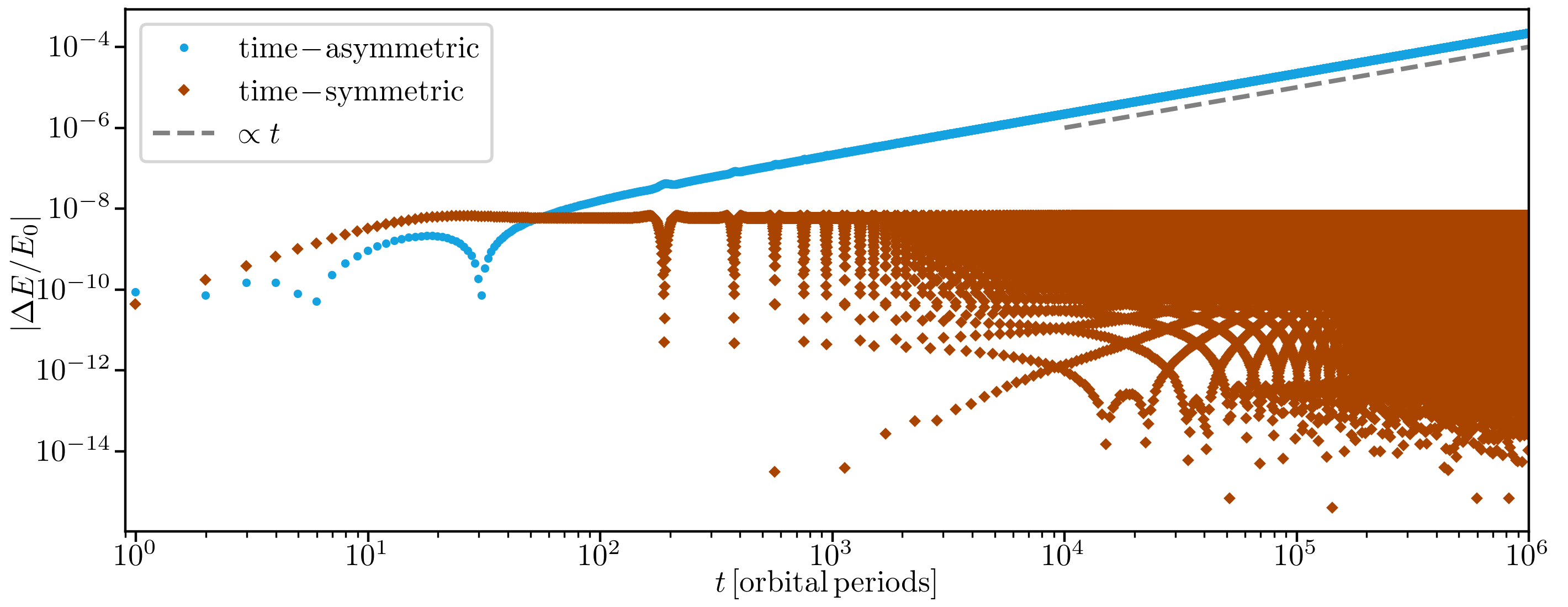}
\caption{The energy error over time during the integration of a Keplerian 2-body system with mass ratio $q=10^{-4}$ and eccentricity $e=0.2$. Energy errors were output every $\delta t=6.25$, leading to aliasing on a timescale of $\sim189$ orbits as the reported energy errors sample different parts of the orbit. The time-symmetric integration schemes results in bounded energy errors over this timeframe, whereas the time-asymmetric scheme accrues energy error linearly in time.}\label{fig:tsym}
\end{figure}

To illustrate the potential benefits of time-symmetric schemes, Figure \ref{fig:tsym} shows the energy errors resulting from the application of both time-symmetric and time-asymmetric 3-point $r=1$ schemes to an unsoftened Keplerian 2-body problem with $e=0.2$ and mass ratio $q=10^{-4}$, using constant timesteps for simplicity and integrating the system for $10^6$ orbital periods. Blue circles display energy errors from the 6th-order integrator of Section \ref{sec:variable}, using the highest-order-available Taylor series predictor and a constant time step of $\Delta t=2^{-4}$. Orange squares mark the energy errors when applying the implicit method of Equation (\ref{eq:sym3x2}) using a time step of $\Delta t=2^{-3}$ such that the inter-node spacing was fixed between integration schemes, iterating each step to a tolerance of $10^{-14}$.\footnote{Note that this method of solution is not strictly necessary, and that predictor-corrector strategies can also achieve time-symmetry, as in \citet{1998MNRAS.297.1067K,2004PASJ...56..861K,2020MNRAS.496.1217D}. Similarly, applying the corrector multiple times to the time-asymmetric scheme here does not improve its secular errors.} In short, the application of a time-asymmetric corrector leads to the accumulation of truncation errors over time, whereas time-symmetric correctors can ameliorate these secular errors. These results apply regardless of whether the corrector is applied in a predictor-corrector scheme or treated implicitly as a system of nonlinear equations. 


\section{The Variable-Timestep 3-Point 12th-Order Scheme}\label{app:3x4}
The 3-point 12th-order generalization of Equation (\ref{eq:herm8}) omitted from Section \ref{sec:variable} is 
{\allowdisplaybreaks
\begin{equation}\label{eq:var34}
\mathbf{v}_1 - \mathbf{v}_0 = \Delta t_1 \sum_{i=-1}^1g_{i,0}\mathbf{a}_i + \Delta t_1^2 \sum_{i=-1}^1g_{i,1}\mathbf{a}^{(1)}_i +\Delta t_1^3 \sum_{i=-1}^2g_{i,1}\mathbf{a}^{(2)}_i+\Delta t_1^4 \sum_{i=-1}^1g_{i,3}\mathbf{a}^{(3)}_i,
\end{equation}
where, defining $\zeta\equiv \Delta t_0 / \Delta t_1$ as illustrated in Figure \ref{fig:schematic1},
\begin{align}
 g_{-1,0} &=& \frac{1}{1386}\frac{363\zeta^6 + 1089\zeta^5 + 1375\zeta^4 + 935\zeta^3 + 363\zeta^2 + 77\zeta +7}{\zeta^7(\zeta^7 + 7\zeta^6 + 21\zeta^5 + 35\zeta^4 + 35\zeta^3 + 21\zeta^2 + 7\zeta +1)}  \\ \nonumber
g_{0,0} &=& \frac{1}{1386}\frac{693\zeta^7 + 176\zeta^6 - 88\zeta^5 + 4\zeta^4 + 38\zeta^3 - 20\zeta^2 - 28\zeta - 7}{\zeta^7}   \\ \nonumber
g_{1,0} &=& \frac{1}{1386}\frac{693\zeta^7+4675\zeta^6+13409\zeta^5+21171\zeta^4+19877\zeta^3+11143\zeta^2 + 3465\zeta +462}{\zeta^7 + 7\zeta^6 + 21\zeta^5 + 35\zeta^4 + 35\zeta^3 + 21\zeta^2 + 7\zeta +1}\\ \nonumber
g_{-1,1} &=& \frac{1}{2772}\frac{198\zeta^5 + 495\zeta^4 + 500\zeta^3 + 255\zeta^2 + 66^\zeta +7}{\zeta^6(\zeta^6 + 6\zeta^5 + 15\zeta^4 + 20\zeta^3 + 15\zeta^2 + 6\zeta +1)}   \\ \nonumber
g_{0,1} &=& \frac{1}{2772}\frac{297\zeta^6 + 176\zeta^5 - 44\zeta^4 - 40\zeta^3 + 34\zeta^2 + 32\zeta +7}{\zeta^6}   \\ \nonumber
g_{1,1} &=& \frac{1}{2772}\frac{-297\zeta^6 - 1606\zeta^5 - 3531\zeta^4 - 4044\zeta^3 - 2585\zeta^2 - 882\zeta - 126}{\zeta^6 + 6\zeta^5 + 15\zeta^4 + 20\zeta^3 + 15\zeta^2 + 6\zeta +1}  \\ \nonumber
g_{-1,2} &=& \frac{1}{13860}\frac{99\zeta^4 + 198\zeta^3 + 152\zeta^2 + 53\zeta +7}{\zeta^5(\zeta^5 + 5\zeta^4 + 10\zeta^3 + 10\zeta^2 + 5\zeta +1)}  \\ \nonumber
g_{0,2} &=& \frac{1}{13860}\frac{165\zeta^5 + 176\zeta^4 + 22\zeta^3 - 62\zeta^2 - 38\zeta - 7}{\zeta^5}  \\ \nonumber
g_{1,2} &=& \frac{1}{13860}\frac{165\zeta^5 + 649\zeta^4 + 968\zeta^3 + 722\zeta^2 + 273\zeta +42}{\zeta^5 + 5\zeta^4 + 10\zeta^3 + 10\zeta^2 + 5\zeta +1}  \\ \nonumber
g_{-1,3} &=& \frac{1}{166320}\frac{44\zeta^3 + 66\zeta^2 + 36\zeta +7}{\zeta^4(\zeta^4 + 4\zeta^3 + 6\zeta^2 + 4\zeta +1)} \\ \nonumber 
g_{0,3} &=& \frac{1}{166320}\frac{99\zeta^4 + 176\zeta^3 + 132\zeta^2 + 48\zeta +7}{\zeta^4(\zeta^4 + 4\zeta^3 + 6\zeta^2 + 4\zeta +1)} \\ \nonumber
g_{1,3} &=& \frac{1}{166320}\frac{-99\zeta^4 - 220\zeta^3 - 198\zeta^2 - 84\zeta - 14}{\zeta^4 + 4\zeta^3 + 6\zeta^2 + 4\zeta +1}. 
\end{align}
}

\subsection{Convergence Tests}\label{app:3x4conv}
\begin{figure}
\includegraphics[width=\linewidth]{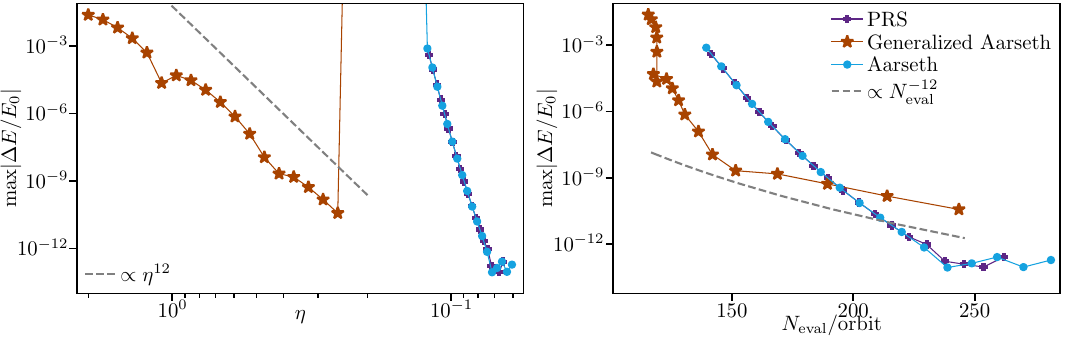}
\caption{Energy errors of the 3-point 12th-order variable-timestep scheme with respect to the dimensionless timestep control $\eta$ and the number of force evaluations for the 2-body test of Section \ref{sec:2bodyvar} (an $e=0.9, q=10^{-4}$ binary). These calculations used a single application of the corrector, Equation (\ref{eq:var34}), each timestep. In this case, only the generalized Aarseth timestepping criterion, Equation (\ref{eq:genaarseth}), results in convergence at the nominal order of the scheme. }\label{fig:3x4}
\end{figure}

\begin{figure}
\includegraphics[width=\linewidth]{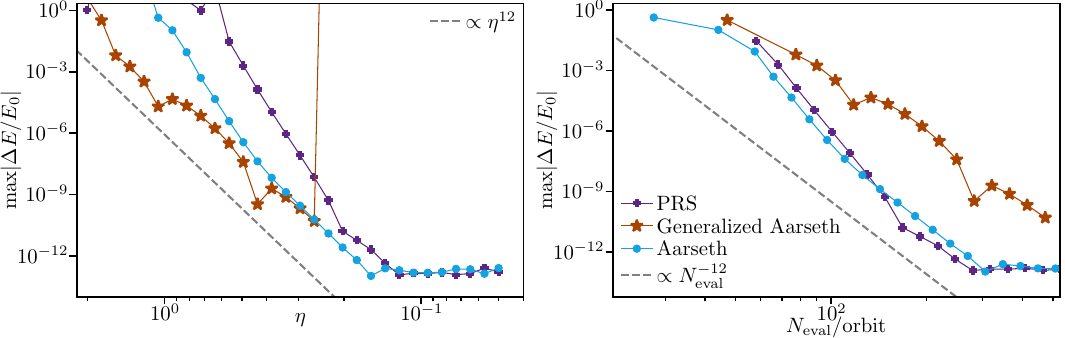}
\caption{Energy errors of the 3-point 12th-order variable-timestep scheme with respect to the dimensionless timestep control $\eta$ and the number of force evaluations for the 2-body test of Section \ref{sec:2bodyvar}. These calculations used two applications of the corrector, Equation (\ref{eq:var34}), each timestep. Although this resulted in convergence at or closer to 12th order, at most error tolerances a single corrector application should be preferred due to the extra cost in force evaluations.}\label{fig:3x4_c2}
\end{figure}

The 3-point 12th-order scheme introduced above is much more sensitive to the timestep employed than the 6th- and 9th-order schemes discussed in Section \ref{sec:variable}.
This is illustrated by the application of the 3-point 12th-order scheme to the eccentric 2-body test of Section \ref{sec:2bodyvar}. Figure \ref{fig:3x4} shows results applying the corrector once each timestep and Figure \ref{fig:3x4_c2} plots results derived applying the corrector twice each time step. At a given dimensionless timestep prefactor $\eta$, applying each corrector expression twice results in far more accurate schemes, possibly by orders of magnitude at a given $\eta$; in terms of the total number of force evaluations however, applying the corrector twice improved efficiency at more lenient error tolerances, but decreased efficiency nearer to machine precision. Although the generalized Aarseth timestepping criterion leads to convegence at the nominal order of the scheme even with only a single corrector iteration, it is susceptible to numerical difficulties thanks to its reliance on very high-order time derivatives of the acceleration and leads to unphysical timesteps as integrations approach machine precision. 
With a single corrector application, the 3-point 12th-order schemes can reach rounding-limited precision more efficiently than the 3-point 9th-order schemes even in spite of the additional floating point operations required to calculate $\mathbf{a}^{(3)}$.

\section{Higher-Order Quadrature-Based Derivatives for Variable-Timestep Schemes}
As discussed in Section \ref{sec:variable}, the quadratures used to construct the correctors for each scheme can be used to approximate higher-order derivatives of the acceleration at the end of the time step. 

\subsection{Derivatives from the 3-point 6th-order scheme}\label{app:3x2derivs}
Using the same underlying quadrature as Equation (\ref{eq:var32}), higher-order derivatives at the end of the time step can be approximated as
\begin{align}\label{eq:c1}
\mathbf{a}^{(k)}_1 \approx \frac{1}{\Delta t_1^k} \sum_{i=-1}^1c^{(k)}_{i,0}\mathbf{a}_i + \frac{1}{\Delta t_1^{k-1}} \sum_{i=-1}^1c^{(k)}_{i,1}\mathbf{a}^{(1)}_i
\end{align}
where $c^{(k)}_{i,j}$ are given by

{\allowdisplaybreaks
\begin{align}\label{eq:c2}
 c^{(2)}_{-1,0} &= 2\frac{5\zeta+2}{\zeta^3(\zeta^2+2\zeta+1)},\, &c^{(2)}_{0,0} =& 2\frac{3\zeta^3+4\zeta^2-\zeta-2}{\zeta^3},\, &c^{(2)}_{1,0} &= 2\frac{-3\zeta^2-10\zeta-10}{\zeta^2+2\zeta+1},\, \\
\nonumber c^{(2)}_{-1,1} &=\frac{2}{\zeta^2(\zeta+1)},\, & c^{(2)}_{0,1} =&2\frac{\zeta^2+2\zeta+1}{\zeta^2},\,&c^{(2)}_{1,1}&=4\frac{\zeta+2}{\zeta+1},\\
\nonumber c^{(3)}_{-1,0} &= 12\frac{5\zeta^2 + 9\zeta +3}{\zeta^3(\zeta^3 + 3\zeta^2 + 3\zeta +1)},\, &c^{(3)}_{0,0} =& 12\frac{\zeta^3 + 4\zeta^2 - 3}{\zeta^3},\, &c^{(3)}_{1,0} &= 12\frac{-\zeta^3 - 7\zeta^2 - 15\zeta - 10}{\zeta^3 + 3\zeta^2 + 3\zeta +1},\, \\
\nonumber c^{(3)}_{-1,1} &= 6\frac{2\zeta +3}{\zeta^2(\zeta^2 + 2\zeta +1)},\, & c^{(3)}_{0,1} =&6\frac{\zeta^2 + 4\zeta +3}{\zeta^2},\,&c^{(3)}_{1,1}&=6\frac{\zeta^2 + 6\zeta +6}{\zeta^2 + 2\zeta +1},\\
\nonumber c^{(4)}_{-1,0} &= 24\frac{-4\zeta^2 + 3\zeta(\zeta +1) + 6(\zeta +1)^2}{\zeta^3(\zeta +1)^3},\, &c^{(4)}_{0,0} =& 24\frac{4\zeta^2 + 3\zeta - 6}{\zeta^3},\, &c^{(4)}_{1,0} &= 24\frac{-4\zeta^2 - 15\zeta - 15}{\zeta^3 + 3\zeta^2 + 3\zeta +1},\, \\
\nonumber c^{(4)}_{-1,1} &=24\frac{\zeta +3}{\zeta^2(\zeta +1)^2},\, & c^{(4)}_{0,1} =&24\frac{2\zeta +3}{\zeta^2},\,&c^{(4)}_{1,1}&=48\frac{\zeta +2}{\zeta^2 + 2\zeta +1},\\
\nonumber c^{(5)}_{-1,0} &= 240\frac{2\zeta+1}{\zeta^3(\zeta+1)^3},\, &c^{(5)}_{0,0} =& 240\frac{\zeta-1}{5\zeta^3},\, &c^{(5)}_{1,0} &= -240\frac{\zeta+2}{\zeta^3+3\zeta^31},\, \\
\nonumber c^{(5)}_{-1,1} &=\frac{120}{\zeta^2(\zeta+1)^2},\, & c^{(5)}_{0,1} =&\frac{120}{\zeta^2},\,&c^{(5)}_{1,1}&=\frac{120}{\zeta^2+2\zeta+1}.
\end{align}
}
Thus, $\mathbf{a},$ and $\mathbf{a}^{(1)}$ are calculated explicitly, and $\mathbf{a}^{(2)}-\mathbf{a}^{(5)}$ can be approximated by taking derivatives of the interpolating polynomial.

\subsection{Derivatives from the 3-point 9th-order scheme}\label{app:3x3derivs}
Using the same underlying quadrature as Equation (\ref{eq:var33}), higher-order derivatives and the end of the time step can be approximated as
\begin{align}
a^{(k)}_2 \approx \frac{1}{\Delta t_1^k} \sum_{i=0}^2c^{(k)}_{i,0}\mathbf{a}_i + \frac{1}{\Delta t_1^{k-1}} \sum_{i=0}^2c^{(k)}_{i,1}\mathbf{a}^{(1)}_i+ \frac{1}{\Delta t_1^{k-2}} \sum_{i=0}^2c^{(k)}_{i,2}\mathbf{a}^{(2)}_i
\end{align}
where $c^{(k)}_{i,j}$ are given by

{\allowdisplaybreaks{\footnotesize
\begin{align}\label{eq:c4}
c^{(3)}_{-1,0} \!&=\!\! 12\frac{14\zeta^2 + 12\zeta +3}{\zeta^5(\zeta^3 + 3\zeta^2 + 3\zeta +1)}, 
&c^{(3)}_{0,0} =& 12\frac{-5\zeta^5 - 9\zeta^4 + 4\zeta^2 - 3\zeta - 3}{\zeta^5} , 
\!\!\!\!&c^{(3)}_{1,0} \!&=\!  12\frac{5\zeta^3 + 24\zeta^2 + 42\zeta +28}{\zeta^3 + 3\zeta^2 + 3\zeta +1} , \\
\nonumber c^{(3)}_{-1,1} \!&=\!\! 6\frac{7\zeta +3}{\zeta^4(\zeta^2 + 2\zeta +1)}  , 
& c^{(3)}_{0,1} =& 6\frac{-4\zeta^4 - 9\zeta^3 - 3\zeta^2 + 5\zeta +3}{\zeta^4}  ,
&c^{(3)}_{1,1}\!&=\! 18\frac{-2\zeta^2 - 7\zeta - 7}{\zeta^2 + 2\zeta +1} ,\\
\nonumber c^{(3)}_{-1,2} \!&=\!\! \frac{3}{\zeta^3(\zeta +1)} , 
& c^{(3)}_{0,2} =& 3\frac{-\zeta^3 - 3\zeta^2 - 3\zeta - 1}{\zeta^3} ,
\!\!\!\!&c^{(3)}_{1,2}\!&=\!  9\frac{\zeta +2}{\zeta +1}  ,\\
\nonumber c^{(4)}_{-1,0} \!&=\!\!  72\frac{28\zeta^3 + 68\zeta^2 + 45\zeta +10}{\zeta^5(\zeta^4 + 4\zeta^3 + 6\zeta^2 + 4\zeta +1)} , 
&c^{(4)}_{0,0} =& 72\frac{-5\zeta^5\!\! -\! 18\zeta^4\! \!-\! 6\zeta^3 \!\!+\! 12\zeta^2\! \!-\! 5\zeta \!-\! 10}{\zeta^5},\! 
\!\!\!\!&c^{(4)}_{1,0} \!&=\! 72\frac{5\zeta^4 + 38\zeta^3 + 108\zeta^2 + 140\zeta +70}{\zeta^4 + 4\zeta^3 + 6\zeta^2 + 4\zeta +1} , \\
\nonumber c^{(4)}_{-1,1} \!&=\!\! 24\frac{21\zeta^2 + 43\zeta +15}{\zeta^4(\zeta^3 + 3\zeta^2 + 3\zeta +1)} , 
& c^{(4)}_{0,1} =& 24\frac{-7\zeta^4\!\! -\! 27\zeta^3\!\! -\! 18\zeta^2\!\! +\! 17\zeta +15}{\zeta^4} ,
\!\!\!\!&c^{(4)}_{1,1}\!&=\! 24\frac{-8\zeta^3 - 51\zeta^2 - 105\zeta - 70}{\zeta^3 + 3\zeta^2 + 3\zeta +1)} ,\\
\nonumber c^{(4)}_{-1,2} \!&=\!\! 12\frac{3\zeta +5}{\zeta^3(\zeta^2 + 2\zeta +1)} , 
& c^{(4)}_{0,2} =& 12\frac{-2\zeta^3\!\! -\! 9\zeta^2\!\! -\! 12\zeta \!-\! 5}{\zeta^3} ,
\!\!\!\!&c^{(4)}_{1,2}\!&=\! 36\frac{\zeta^2 + 5\zeta +5}{\zeta^2 + 2\zeta +1} ,\\
\nonumber c^{(5)}_{-1,0} \!&=\!\! 720\frac{14\zeta^4 + 64\zeta^3 + 91\zeta^2 + 50\zeta +10}{\zeta^5(\zeta^5 + 5\zeta^4 + 10\zeta^3 + 10\zeta^2 + 5\zeta +1)} , 
&c^{(5)}_{0,0} =& 720\frac{-\zeta^5 - 9\zeta^4 - 9\zeta^3 + 9\zeta^2 - 10}{\zeta^5} , 
\!\!\!\!&c^{(5)}_{1,0} \!&=\! 720\frac{\zeta^5\!\! +\! 14\zeta^4\!\! +\! 64\zeta^3\!\! + \!136\zeta^2\!\! +\! 140\zeta\! +\! 56}{\zeta^5 + 5\zeta^4 + 10\zeta^3 + 10\zeta^2 + 5\zeta +1} , \\
\nonumber c^{(5)}_{-1,1} \!&=\!\! 360\frac{7\zeta^3 + 30\zeta^2 + 34\zeta +10}{\zeta^4(\zeta^4 + 4\zeta^3 + 6\zeta^2 + 4\zeta +1)} , 
& c^{(5)}_{0,1} =& 360\frac{-\zeta^4 - 9\zeta^3 - 12\zeta^2 + 6\zeta +10}{\zeta^4} ,
\!\!\!\!&c^{(5)}_{1,1}\!&=\! 360\frac{-\zeta^4 - 13\zeta^3 - 48\zeta^2 - 70\zeta - 35}{\zeta^4 + 4\zeta^3 + 6\zeta^2 + 4\zeta +1} ,\\
\nonumber c^{(5)}_{-1,2} \!&=\!\! 60\frac{3\zeta^2 + 12\zeta +10}{\zeta^3(\zeta^3 + 3\zeta^2 + 3\zeta +1)} , 
& c^{(5)}_{0,2} =& 60\frac{-\zeta^3 - 9\zeta^2 - 18\zeta - 10}{\zeta^3} ,
\!\!\!\!&c^{(5)}_{1,2}\!&=\! 60\frac{\zeta^3 + 12\zeta^2 + 30\zeta +20}{(\zeta^3 + 3\zeta^2 + 3\zeta +1)} ,\\
\nonumber c^{(6)}_{-1,0} \!&=\!\! 1440\frac{14\zeta^4 + 112\zeta^3 + 212\zeta^2 + 135\zeta +30}{\zeta^5(\zeta^5 + 5\zeta^4 + 10\zeta^3 + 10\zeta^2 + 5\zeta +1)} , 
&c^{(6)}_{0,0} =& 1440\frac{-9\zeta^4\!\! -\! 27\zeta^3\!\! +\! 13\zeta^2\!\! +\! 15\zeta \!-\! 30}{\zeta^5} , 
\!\!\!\!&c^{(6)}_{1,0} \!&=\! 1440\frac{9\zeta^4\!\! +\! 72\zeta^3\!\! +\! 212\zeta^2\!\! +\! 280\zeta \!+ \!140}{(\zeta^5\!\! + \!5\zeta^4\!\! +\! 10\zeta^3\!\! +\! 10\zeta^2\!\! + \!5\zeta +1)} , \\
\nonumber c^{(6)}_{-1,1} \!&=\!\! 720\frac{7\zeta^3 + 56\zeta^2 + 88\zeta +30}{\zeta^4(\zeta^4 + 4\zeta^3 + 6\zeta^2 + 4\zeta +1)} , 
& c^{(6)}_{0,1} =& 720\frac{-9\zeta^3 - 30\zeta^2 + 2\zeta +30}{\zeta^4} ,
\!\!\!\!&c^{(6)}_{1,1}\!&=\! 2160\frac{-3\zeta^3 - 20\zeta^2 - 42\zeta - 28}{\zeta^4 + 4\zeta^3 + 6\zeta^2 + 4\zeta +1)},\\
\nonumber c^{(6)}_{-1,2} \!&=\!\!360\frac{3\zeta^2 - 12\zeta(\zeta +1) + 10(\zeta +1)^2}{\zeta^3(\zeta +1)^3} , 
& c^{(6)}_{0,2} =& 360\frac{-3\zeta^2 - 12\zeta - 10}{\zeta^3} ,
\!\!\!\!&c^{(6)}_{1,2}\!&=\! 1080\frac{\zeta^2 + 5\zeta +5}{\zeta^3 + 3\zeta^2 + 3\zeta +1} ,\\
\nonumber c^{(7)}_{-1,0} \!&=\!\! 30240\frac{-3\zeta^3\!\!\!+\!\! \zeta^2(\zeta\!\! +\! 1)\! +\! 5\zeta(\zeta\!\!+\! 1)^2\! \!+\! 5(\zeta \!\!+\! 1)^3}{\zeta^5(\zeta +1)^5} , 
&c^{(7)}_{0,0} =& 30240\frac{-3\zeta^3 - \zeta^2 + 5\zeta - 5}{\zeta^5} , 
\!\!\!\!&c^{(7)}_{1,0} \!&=\! 30240\frac{3\zeta^3\!\! +\! 16\zeta^2 \!\!+ \!30\zeta \!+\! 20}{\zeta^5\!\! + \!5\zeta^4 \!\!+ \!10\zeta^3 \!\!+ \!10\zeta^2 \!+\! 5\zeta +1} , \\
\nonumber c^{(7)}_{-1,1} \!&=\!\! 5040\frac{-9\zeta^2 + 7\zeta(\zeta +1) + 15(\zeta +1)^2}{\zeta^4(\zeta +1)^4} , 
& c^{(7)}_{0,1} =& 5040\frac{-9\zeta^2 - 7\zeta +15}{\zeta^4} ,
\!\!\!\!&c^{(7)}_{1,1}\!&=\! 5040\frac{-9\zeta^2 - 35\zeta - 35}{(\zeta^4 + 4\zeta^3 + 6\zeta^2 + 4\zeta +1} ,\\
\nonumber c^{(7)}_{-1,2} \!&=\!\! 2520\frac{2\zeta +5}{\zeta^3(\zeta +1)^3} , 
& c^{(7)}_{0,2} =& 2520\frac{-3\zeta - 5}{\zeta^3} ,
\!\!\!\!&c^{(7)}_{1,2}\!&=\! 7560\frac{\zeta +2}{\zeta^3 + 3\zeta^2 + 3\zeta +1} ,\\
\nonumber c^{(8)}_{-1,0} \!&=\!\!  120960\frac{2\zeta^2 + 3\zeta(\zeta +1) + 2(\zeta +1)^2}{\zeta^5(\zeta +1)^5}, 
&c^{(8)}_{0,0} =& 120960\frac{2\zeta +1}{\zeta^4(\zeta +1)64} , 
\!\!\!\!&c^{(8)}_{1,0} \!&=\! \frac{20160}{\zeta^3(\zeta +1)^3} , \\
\nonumber c^{(8)}_{-1,1} \!&=\!\! 120960\frac{-2\zeta^2 + 3\zeta - 2}{\zeta^5} , 
& c^{(8)}_{0,1} =& 120960\frac{1 - \zeta}{\zeta^4} ,
\!\!\!\!&c^{(8)}_{1,1}\!&=\! -\frac{20160}{\zeta^3} ,\\
\nonumber c^{(8)}_{-1,2} \!&=\!\! 120960\frac{2\zeta^2 + 7\zeta +7}{\zeta^5 + 5\zeta^4 + 10\zeta^3 + 10\zeta^2 + 5\zeta +1} , 
& c^{(8)}_{0,2} =& 120960\frac{-\zeta - 2}{\zeta^4 + 4\zeta^3 + 6\zeta^2 + 4\zeta +1} ,
\!\!\!\!&c^{(8)}_{1,2}\!&=\! \frac{20160}{\zeta^3 + 3\zeta^2 + 3\zeta +1}.
\end{align}
}}
Thus, $\mathbf{a}-\mathbf{a}^{(2)}$ are calculated explicitly, and $\mathbf{a}^{(3)}-\mathbf{a}^{(8)}$ can be approximated by taking derivatives of the interpolating polynomial.


\subsection{Derivatives from the 3-point 12th-order scheme}\label{app:3x4derivs}
Using the same underlying quadrature as Equation (\ref{eq:var34}), higher-order derivatives and the end of the time step can be approximated as
\begin{align}
a^{(k)}_2 \approx \frac{1}{\Delta t_1^k} \sum_{i=0}^2c^{(k)}_{i,0}\mathbf{a}_i + \frac{1}{\Delta t_1^{k-1}} \sum_{i=0}^2c^{(k)}_{i,1}\mathbf{a}^{(1)}_i+ \frac{1}{\Delta t_1^{k-2}} \sum_{i=0}^2c^{(k)}_{i,2}\mathbf{a}^{(2)}_i
+ \frac{1}{\Delta t_1^{k-3}} \sum_{i=0}^2c^{(k)}_{i,3}\mathbf{a}^{(3)}_i
\end{align}
where $c^{(k)}_{i,j}$ are given by

\allowdisplaybreaks
\begin{align}\label{eq:c6}
c^{(4)}_{-1,0} \!&=\!\! 120\frac{33\zeta^3 + 44\zeta^2 + 22\zeta +4}{\zeta^7(\zeta^4 + 4\zeta^3 + 6\zeta^2 + 4\zeta +1)}, \\
\nonumber c^{(4)}_{0,0} &=      120\frac{7\zeta^7 + 16\zeta^6 + 4\zeta^5 - 8\zeta^4 + 3\zeta^3 + 4\zeta^2 - 6\zeta - 4}{\zeta^7}, \\
\nonumber c^{(4)}_{1,0} \!&=\! 120\frac{-7\zeta^4 - 44\zeta^3 - 110\zeta^2 - 132\zeta - 66}{\zeta^4 + 4\zeta^3 + 6\zeta^2 + 4\zeta +1} , \\
\nonumber c^{(4)}_{-1,1} \!&=\!\! 120\frac{9\zeta^2 + 8\zeta +2}{\zeta^6(\zeta^3 + 3\zeta^2 + 3\zeta +1)}, \\
\nonumber c^{(4)}_{0,1} &=      120\frac{3\zeta^6 + 8\zeta^5 + 4\zeta^4 - 4\zeta^3 - \zeta^2 + 4\zeta +2}{\zeta^6}, \\
\nonumber c^{(4)}_{1,1} \!&=\! 480\frac{\zeta^3 + 5\zeta^2 + 9\zeta +6}{\zeta^3 + 3\zeta^2 + 3\zeta +1} , \\
\nonumber c^{(4)}_{-1,2} \!&=\!\! 12\frac{9\zeta +4}{\zeta^5(\zeta^2 + 2\zeta +1)}, \\
\nonumber c^{(4)}_{0,2} &=      12\frac{5\zeta^5 + 16\zeta^4 + 14\zeta^3 - 4\zeta^2 - 11\zeta - 4}{\zeta^5}, \\
\nonumber c^{(4)}_{1,2} \!&=\! 24\frac{-5\zeta^2 - 18\zeta - 18}{\zeta^2 + 2\zeta +1} , \\
\nonumber c^{(4)}_{-1,3} \!&=\!\! \frac{4}{\zeta^4(\zeta +1)}, \\
\nonumber c^{(4)}_{0,3} &=      4\frac{\zeta^4 + 4\zeta^3 + 6\zeta^2 + 4\zeta +1}{\zeta^4}, \\
\nonumber c^{(4)}_{1,3} \!&=\! 16\frac{\zeta +2}{\zeta +1} , \\
\nonumber c^{(5)}_{-1,0} \!&=\!\! 480\frac{165\zeta^4 + 495\zeta^3 + 484\zeta^2 + 210\zeta +35}{\zeta^7(\zeta^5 + 5\zeta^4 + 10\zeta^3 + 10\zeta^2 + 5\zeta +1)}, \\
\nonumber c^{(5)}_{0,0} &=      480\frac{21\zeta^7 + 80\zeta^6 + 50\zeta^5 - 50\zeta^4 + 41\zeta^2 - 35\zeta - 35}{5\zeta^7}, \\
\nonumber c^{(5)}_{1,0} \!&=\! 480\frac{-21\zeta^5 - 185\zeta^4 - 660\zeta^3 - 1210\zeta^2 - 1155\zeta - 462}{\zeta^5 + 5\zeta^4 + 10\zeta^3 + 10\zeta^2 + 5\zeta +1}, \\
\nonumber c^{(5)}_{-1,1} \!&=\!\! 120\frac{180\zeta^3 + 465\zeta^2 + 316\zeta +70}{\zeta^6(\zeta^4 + 4\zeta^3 + 6\zeta^2 + 4\zeta +1)}, \\
\nonumber c^{(5)}_{0,1} &=     120\frac{39\zeta^6 + 160\zeta^5 + 140\zeta^4 - 80\zeta^3 - 65\zeta^2 + 104\zeta +70}{\zeta^6}, \\
\nonumber c^{(5)}_{1,1} \!&=\! 600\frac{9\zeta^4 + 68\zeta^3 + 194\zeta^2 + 252\zeta +126}{\zeta^4 + 4\zeta^3 + 6\zeta^2 + 4\zeta +1} , \\
\nonumber c^{(5)}_{-1,2} \!&=\!\! 120\frac{18\zeta^2 + 39\zeta +14}{\zeta^5(\zeta^3 + 3\zeta^2 + 3\zeta +1)}, \\
\nonumber c^{(5)}_{0,2} &=      120\frac{7\zeta^5 + 32\zeta^4 + 40\zeta^3 - 2\zeta^2 - 31\zeta - 14}{\zeta^5}, \\
\nonumber c^{(5)}_{1,2} \!&=\! 240\frac{-5\zeta^3 - 31\zeta^2 - 63\zeta - 42}{\zeta^3 + 3\zeta^2 + 3\zeta +1} , \\
\nonumber c^{(5)}_{-1,3} \!&=\!\! 20\frac{4\zeta +7}{\zeta^4(\zeta^2 + 2\zeta +1)}, \\
\nonumber c^{(5)}_{0,3} &=      20\frac{3\zeta^4 + 16\zeta^3 + 30\zeta^2 + 24\zeta +7}{\zeta^4}, \\
\nonumber c^{(5)}_{1,3} \!&=\! 40\frac{3\zeta^2 + 14\zeta +14}{\zeta^2 + 2\zeta +1} , \\
\nonumber c^{(6)}_{-1,0} \!&=\!\! 7200\frac{99\zeta^5 + 506\zeta^4 + 891\zeta^3 + 708\zeta^2 + 273\zeta +42}{\zeta^7(\zeta^6 + 6\zeta^5 + 15\zeta^4 + 20\zeta^3 + 15\zeta^2 + 6\zeta +1)}, \\
\nonumber c^{(6)}_{0,0} &=      7200\frac{7\zeta^7 + 48\zeta^6 + 60\zeta^5 - 32\zeta^4 - 24\zeta^3 + 48\zeta^2 - 21\zeta - 42}{\zeta^7}, \\
\nonumber c^{(6)}_{1,0} \!&=\! 7200\frac{-7\zeta^6 - 90\zeta^5 - 453\zeta^4 - 1188\zeta^3 - 1749\zeta^2 - 1386\zeta - 462}{(\zeta^6 + 6\zeta^5 + 15\zeta^4 + 20\zeta^3 + 15\zeta^2 + 6\zeta +1} , \\
\nonumber c^{(6)}_{-1,1} \!&=\!\! 1440\frac{135\zeta^4 + 640\zeta^3 + 945\zeta^2 + 528\zeta +105}{\zeta^6(\zeta^5 + 5\zeta^4 + 10\zeta^3 + 10\zeta^2 + 5\zeta +1)}, \\
\nonumber c^{(6)}_{0,1} &=      1440\frac{17\zeta^6 + 120\zeta^5 + 180\zeta^4 - 40\zeta^3 - 120\zeta^2 + 102\zeta +105}{\zeta^6}, \\
\nonumber c^{(6)}_{1,1} \!&=\! 8640\frac{3\zeta^5 + 35\zeta^4 + 150\zeta^3 + 310\zeta^2 + 315\zeta +126}{\zeta^5 + 5\zeta^4 + 10\zeta^3 + 10\zeta^2 + 5\zeta +1} , \\
\nonumber c^{(6)}_{-1,2} \!&=\!\! 360\frac{54\zeta^3 + 236\zeta^2 + 279\zeta +84}{\zeta^5(\zeta^4 + 4\zeta^3 + 6\zeta^2 + 4\zeta +1)}, \\
\nonumber c^{(6)}_{0,2} &=     360\frac{13\zeta^5 + 96\zeta^4 + 180\zeta^3 + 40\zeta^2 - 141\zeta - 84}{\zeta^5}, \\
\nonumber c^{(6)}_{1,2} \!&=\! 1080\frac{-5\zeta^4 - 52\zeta^3 - 178\zeta^2 - 252\zeta - 126}{\zeta^4 + 4\zeta^3 + 6\zeta^2 + 4\zeta +1} , \\
\nonumber c^{(6)}_{-1,3} \!&=\!\! 360\frac{2\zeta^2 + 8\zeta +7}{\zeta^4(\zeta^3 + 3\zeta^2 + 3\zeta +1)}, \\
\nonumber c^{(6)}_{0,3} &=      360\frac{\zeta^4 + 8\zeta^3 + 20\zeta^2 + 20\zeta +7}{\zeta^4}, \\
\nonumber c^{(6)}_{1,3} \!&=\! 480\frac{\zeta^3 + 9\zeta^2 + 21\zeta +14}{\zeta^3 + 3\zeta^2 + 3\zeta +1} , \\
\nonumber c^{(7)}_{-1,0} \!&=\!\! 100800\frac{33\zeta^6 + 275\zeta^5 + 781\zeta^4 + 1033\zeta^3 + 703\zeta^2 + 245\zeta +35}{\zeta^7(\zeta^7 + 7\zeta^6 + 21\zeta^5 + 35\zeta^4 + 35\zeta^3 + 21\zeta^2 + 7\zeta +1)}, \\
\nonumber c^{(7)}_{0,0} &=      100800\frac{\zeta^7 + 16\zeta^6 + 40\zeta^5 - 4\zeta^4 - 32\zeta^3 + 32\zeta^2 - 35}{\zeta^7}, \\
\nonumber c^{(7)}_{1,0} \!&=\! 100800\frac{-\zeta^7 - 23\zeta^6 - 173\zeta^5 - 647\zeta^4 - 1375\zeta^3 - 1705\zeta^2 - 1155\zeta - 330}{(\zeta^7 + 7\zeta^6 + 21\zeta^5 + 35\zeta^4 + 35\zeta^3 + 21\zeta^2 + 7\zeta +1},\\
\nonumber c^{(7)}_{-1,1} \!&=\!\! 50400\frac{18\zeta^5 + 145\zeta^4 + 372\zeta^3 + 405\zeta^2 + 194\zeta +35}{\zeta^6(\zeta^6 + 6\zeta^5 + 15\zeta^4 + 20\zeta^3 + 15\zeta^2 + 6\zeta +1}, \\
\nonumber c^{(7)}_{0,1} &=      50400\frac{\zeta^6 + 16\zeta^5 + 44\zeta^4 + 8\zeta^3 - 40\zeta^2 + 16\zeta +35}{\zeta^6}, \\
\nonumber c^{(7)}_{1,1} \!&=\! 50400\frac{\zeta^6 + 22\zeta^5 + 147\zeta^4 + 460\zeta^3 + 755\zeta^2 + 630\zeta +210}{\zeta^6 + 6\zeta^5 + 15\zeta^4 + 20\zeta^3 + 15\zeta^2 + 6\zeta +1} , \\
\nonumber c^{(7)}_{-1,2} \!&=\!\! 10080\frac{9\zeta^4 + 70\zeta^3 + 160\zeta^2 + 135\zeta +35}{\zeta^5(\zeta^5 + 5\zeta^4 + 10\zeta^3 + 10\zeta^2 + 5\zeta +1}, \\
\nonumber c^{(7)}_{0,2} &=      10080\frac{\zeta^5 + 16\zeta^4 + 50\zeta^3 + 30\zeta^2 - 40\zeta - 35}{\zeta^5}, \\
\nonumber c^{(7)}_{1,2} \!&=\! 10080\frac{-\zeta^5 - 21\zeta^4 - 120\zeta^3 - 290\zeta^2 - 315\zeta - 126}{\zeta^5 + 5\zeta^4 + 10\zeta^3 + 10\zeta^2 + 5\zeta +1}, \\
\nonumber c^{(7)}_{-1,3} \!&=\!\! 840\frac{4\zeta^3 + 30\zeta^2 + 60\zeta +35}{\zeta^4(\zeta^4 + 4\zeta^3 + 6\zeta^2 + 4\zeta +1}, \\
\nonumber c^{(7)}_{0,3} &=      840\frac{\zeta^4 + 16\zeta^3 + 60\zeta^2 + 80\zeta +35}{\zeta^4}, \\
\nonumber c^{(7)}_{1,3} \!&=\! 840\frac{\zeta^4 + 20\zeta^3 + 90\zeta^2 + 140\zeta +70}{\zeta^4 + 4\zeta^3 + 6\zeta^2 + 4\zeta +1} , \\
\nonumber c^{(8)}_{-1,0} \!&=\!\! 201600\frac{-16\zeta^6 + 100\zeta^5(\zeta +1) - 56\zeta^4(\zeta +1)^2 - 137\zeta^3(\zeta +1)^3 - 68\zeta^2(\zeta +1)^4 + 70\zeta(\zeta +1)^5 + 140(\zeta +1)^6}{\zeta^7(\zeta +1)^7}, \\
\nonumber c^{(8)}_{0,0} &=     201600\frac{16\zeta^6 + 100\zeta^5 + 56\zeta^4 - 137\zeta^3 + 68\zeta^2 + 70\zeta - 140}{\zeta^7}, \\
\nonumber c^{(8)}_{1,0} \!&=\! 201600\frac{-16\zeta^6 - 212\zeta^5 - 1092\zeta^4 - 2915\zeta^3 - 4345\zeta^2 - 3465\zeta - 1155}{\zeta^7 + 7\zeta^6 + 21\zeta^5 + 35\zeta^4 + 35\zeta^3 + 21\zeta^2 + 7\zeta +1} , \\
\nonumber c^{(8)}_{-1,1} \!&=\!\! 201600\frac{-8\zeta^5 + 52\zeta^4(\zeta +1) - 44\zeta^3(\zeta +1)^2 - 65\zeta^2(\zeta +1)^3 + 4\zeta(\zeta +1)^4 + 70(\zeta +1)^5}{\zeta^6(\zeta +1)^6}, \\
\nonumber c^{(8)}_{0,1} &=     201600\frac{8\zeta^5 + 52\zeta^4 + 44\zeta^3 - 65\zeta^2 - 4\zeta +70}{\zeta^6}, \\
\nonumber c^{(8)}_{1,1} \!&=\! 806400\frac{2\zeta^5 + 24\zeta^4 + 105\zeta^3 + 220\zeta^2 + 225\zeta +90}{\zeta^6 + 6\zeta^5 + 15\zeta^4 + 20\zeta^3 + 15\zeta^2 + 6\zeta +1} , \\
\nonumber c^{(8)}_{-1,2} \!&=\!\! 20160\frac{-16\zeta^4 + 110\zeta^3(\zeta +1) - 140\zeta^2(\zeta +1)^2 - 85\zeta(\zeta +1)^3 + 140(\zeta +1)^4}{\zeta^5(\zeta +1)^5}, \\
\nonumber c^{(8)}_{0,2} &=      20160\frac{16\zeta^4 + 110\zeta^3 + 140\zeta^2 - 85\zeta - 140}{\zeta^5}, \\
\nonumber c^{(8)}_{1,2} \!&=\! 40320\frac{-8\zeta^4 - 85\zeta^3 - 295\zeta^2 - 420\zeta - 210}{(\zeta^5 + 5\zeta^4 + 10\zeta^3 + 10\zeta^2 + 5\zeta +1} , \\
\nonumber c^{(8)}_{-1,3} \!&=\!\! 6720\frac{-4\zeta^3 + 30\zeta^2(\zeta +1) - 60\zeta(\zeta +1)^2 + 35(\zeta +1)^3}{\zeta^4(\zeta +1)^4}, \\
\nonumber c^{(8)}_{0,3} &=      6720\frac{4\zeta^3 + 30\zeta^2 + 60\zeta +35}{\zeta^4}, \\
\nonumber c^{(8)}_{1,3} \!&=\! 26880\frac{\zeta^3 + 9\zeta^2 + 21\zeta +14}{\zeta^4 + 4\zeta^3 + 6\zeta^2 + 4\zeta +1} , \\
\nonumber c^{(9)}_{-1,0} \!&=\!\! 7257600\frac{6\zeta^5 - 14\zeta^4(\zeta +1) - 15\zeta^3(\zeta +1)^2 + 3\zeta^2(\zeta +1)^3 + 21\zeta(\zeta +1)^4 + 21(\zeta +1)^5}{\zeta^7(\zeta +1)^7}, \\
\nonumber c^{(9)}_{0,0} &=      7257600\frac{6\zeta^5 + 14\zeta^4 - 15\zeta^3 - 3\zeta^2 + 21\zeta - 21}{\zeta^7}, \\
\nonumber c^{(9)}_{1,0} \!&=\! 7257600\frac{-6\zeta^5 - 56\zeta^4 - 209\zeta^3 - 396\zeta^2 - 385\zeta - 154}{\zeta^7 + 7\zeta^6 + 21\zeta^5 + 35\zeta^4 + 35\zeta^3 + 21\zeta^2 + 7\zeta +1}, \\
\nonumber c^{(9)}_{-1,1} \!&=\!\! 1814400\frac{12\zeta^4 - 32\zeta^3(\zeta +1) - 21\zeta^2(\zeta +1)^2 + 24\zeta(\zeta +1)^3 + 42(\zeta +1)^4}{\zeta^6(\zeta +1)^6}, \\
\nonumber c^{(9)}_{0,1} &=      1814400\frac{12\zeta^4 + 32\zeta^3 - 21\zeta^2 - 24\zeta +42}{\zeta^6}, \\
\nonumber c^{(9)}_{1,1} \!&=\! 5443200\frac{4\zeta^4 + 32\zeta^3 + 95\zeta^2 + 126\zeta +63}{\zeta^6 + 6\zeta^5 + 15\zeta^4 + 20\zeta^3 + 15\zeta^2 + 6\zeta +1} , \\
\nonumber c^{(9)}_{-1,2} \!&=\!\! 362880\frac{12\zeta^3 - 38\zeta^2(\zeta +1) - 3\zeta(\zeta +1)^2 + 42(\zeta +1)^3}{\zeta^5(\zeta +1)^5}, \\
\nonumber c^{(9)}_{0,2} &=     362880\frac{12\zeta^3 + 38\zeta^2 - 3\zeta - 42}{\zeta^5}, \\
\nonumber c^{(9)}_{1,2} \!&=\! 2177280\frac{-2\zeta^3 - 13\zeta^2 - 27\zeta - 18}{(\zeta^5 + 5\zeta^4 + 10\zeta^3 + 10\zeta^2 + 5\zeta +1} , \\
\nonumber c^{(9)}_{-1,3} \!&=\!\! 181440\frac{2\zeta^2 - 8\zeta(\zeta +1) + 7(\zeta +1)^2}{\zeta^4(\zeta +1)^4}, \\
\nonumber c^{(9)}_{0,3} &=      181440\frac{2\zeta^2 + 8\zeta +7}{\zeta^4}, \\
\nonumber c^{(9)}_{1,3} \!&=\! 120960\frac{3\zeta^2 + 14\zeta +14}{\zeta^4 + 4\zeta^3 + 6\zeta^2 + 4\zeta +1} , \\
\nonumber c^{(10)}_{-1,0} \!&=\!\! 7257600\frac{-40\zeta^4 - 5\zeta^3(\zeta +1) + 68\zeta^2(\zeta +1)^2 + 105\zeta(\zeta +1)^3 + 70(\zeta +1)^4}{\zeta^7(\zeta +1)^7}, \\
\nonumber c^{(10)}_{0,0} &=      7257600\frac{40\zeta^4 - 5\zeta^3 - 68\zeta^2 + 105\zeta - 70}{\zeta^7}, \\
\nonumber c^{(10)}_{1,0} \!&=\! 7257600\frac{-40\zeta^4 - 275\zeta^3 - 737\zeta^2 - 924\zeta - 462}{\zeta^7 + 7\zeta^6 + 21\zeta^5 + 35\zeta^4 + 35\zeta^3 + 21\zeta^2 + 7\zeta +1} , \\
\nonumber c^{(10)}_{-1,1} \!&=\!\! 7257600\frac{-20\zeta^3 + 5\zeta^2(\zeta +1) + 38\zeta(\zeta +1)^2 + 35(\zeta +1)^3}{\zeta^6(\zeta +1)^6}, \\
\nonumber c^{(10)}_{0,1} &=      7257600\frac{20\zeta^3 + 5\zeta^2 - 38\zeta +35}{\zeta^6}, \\
\nonumber c^{(10)}_{1,1} \!&=\! 72576000\frac{2\zeta^3 + 11\zeta^2 + 21\zeta +14}{\zeta^6 + 6\zeta^5 + 15\zeta^4 + 20\zeta^3 + 15\zeta^2 + 6\zeta +1} , \\
\nonumber c^{(10)}_{-1,2} \!&=\!\! 1814400\frac{-16\zeta^2 + 13\zeta(\zeta +1) + 28(\zeta +1)^2}{\zeta^5(\zeta +1)^5}, \\
\nonumber c^{(10)}_{0,2} &=      1814400\frac{16\zeta^2 + 13\zeta - 28}{\zeta^5}, \\ 
\nonumber c^{(10)}_{1,2} \!&=\! 1814400\frac{-16\zeta^2 - 63\zeta - 63}{\zeta^5 + 5\zeta^4 + 10\zeta^3 + 10\zeta^2 + 5\zeta +1} , \\
\nonumber c^{(10)}_{-1,3} \!&=\!\!  604800\frac{3\zeta +7}{\zeta^4(\zeta +1)^4}, \\
\nonumber c^{(10)}_{0,3} &=      604800\frac{4\zeta +7}{\zeta^4}, \\
\nonumber c^{(10)}_{1,3} \!&=\! 2419200\frac{\zeta +2}{\zeta^4 + 4\zeta^3 + 6\zeta^2 + 4\zeta +1} , \\
\nonumber c^{(11)}_{-1,0} \!&=\!\! 798336000\frac{\zeta 3 + 2\zeta^2(\zeta +1) + 2\zeta(\zeta +1)^2 + (\zeta +1)^3}{\zeta^7(\zeta +1)^7}, \\
\nonumber c^{(11)}_{0,0} &=     798336000\frac{\zeta^3 - 2\zeta^2 + 2\zeta - 1}{\zeta^7}, \\
\nonumber c^{(11)}_{1,0} \!&=\! 798336000\frac{-\zeta^3 - 5\zeta^2 - 9\zeta - 6}{\zeta^7 + 7\zeta^6 + 21\zeta^5 + 35\zeta^4 + 35\zeta^3 + 21\zeta^2 + 7\zeta +1}, \\
\nonumber c^{(11)}_{-1,1} \!&=\!\! 79833600\frac{5\zeta^2 + 8\zeta(\zeta +1) + 5(\zeta +1)^2}{\zeta^6(\zeta +1)^6}, \\
\nonumber c^{(11)}_{0,1} &=      79833600\frac{5\zeta^2 - 8\zeta +5}{\zeta 6}, \\
\nonumber c^{(11)}_{1,1} \!&=\! 79833600\frac{5\zeta^2 + 18\zeta +18}{\zeta^6 + 6\zeta^5 + 15\zeta^4 + 20\zeta^3 + 15\zeta^2 + 6\zeta +1}, \\
\nonumber c^{(11)}_{-1,2} \!&=\!\! 79833600\frac{2\zeta +1}{\zeta^5(\zeta +1)^5}, \\
\nonumber c^{(11)}_{0,2} &=      79833600\frac{\zeta - 1}{\zeta^5}, \\
\nonumber c^{(11)}_{1,2} \!&=\! 79833600\frac{-\zeta - 2}{\zeta^5 + 5\zeta^4 + 10\zeta^3 + 10\zeta^2 + 5\zeta +1} , \\
\nonumber c^{(11)}_{-1,3} \!&=\!\! \frac{6652800}{\zeta^4(\zeta +1)^4}, \\
\nonumber c^{(11)}_{0,3} &=      \frac{6652800}{\zeta^4}, \\
\nonumber c^{(11)}_{1,3} \!&=\! \frac{6652800}{\zeta^4 + 4\zeta^3 + 6\zeta^2 + 4\zeta +1} .
\end{align}
Thus, $\mathbf{a}-\mathbf{a}^{(3)}$ are calculated explicitly, and $\mathbf{a}^{(4)}-\mathbf{a}^{(11)}$ can be approximated by taking derivatives of the interpolating polynomial.

\section{Additional Time-Symmetric Quadratures}\label{app:timesym}
Section \ref{sec:collocation} presented a 3-point 2-derivative quadrature as an example. The remainder of this section lists the multi-point quadratures omitted from Section \ref{sec:collocation} that were tested in Figure \ref{fig:alltimesym}.

\subsection{The 3-point 3-derivative Quadrature}
The integration scheme using three equally-spaced points and three derivatives at each point is 
\begin{align}
\mathbf{v}_{1/2}&=\mathbf{v}_0 + \frac{\Delta t}{26880}\left(5669\mathbf{a}_0 + 8192\mathbf{a}_{1/2} - 421\mathbf{a}_1\right) +\frac{\Delta t^2}{17920}\left(303\mathbf{a}_0^{(1)} - 560\mathbf{a}_{1/2}^{(1)} + 47\mathbf{a}_1^{(1)}\right) \\ \nonumber &+ \frac{\Delta t^3}{322560}\left( 169\mathbf{a}_{0}^{(2)} + 1024\mathbf{a}_{1/2}^{(2)} -41\mathbf{a}_{1}^{(2)} \right)
\\
\mathbf{v}_{1} &= \mathbf{v}_0 + \frac{\Delta t}{210}\left(41\mathbf{a}_0 + 128\mathbf{a}_{1/2} + 41\mathbf{a}_1 \right) + \frac{\Delta t^2}{70}\left(\mathbf{a}_0^{(1)}-\mathbf{a}_1^{(1)}\right) + \frac{\Delta t^3}{2520}\left( \mathbf{a}_{0}^{(2)} + 16\mathbf{a}_{1/2}^{(2)} + \mathbf{a}_{1}^{(2)} \right).
\end{align}
Similarly to Simpson's rule, in this case three equally-spaced points produce a Gaussian-type quadrature, and this scheme achieves 10th-order accuracy rather than 9th. 

\subsection{Four-point Quadratures}
First, the integration scheme using four equally-spaced points and both the acceleration at each point and its first derivative is
\begin{align}
\mathbf{v}_{1/3} &= \mathbf{v}_0 + \frac{\Delta t}{54432}\left(6893\mathbf{a}_0 + 8451\mathbf{a}_{1/3} + 2403\mathbf{a}_{2/3} + 397\mathbf{a}_1\right) \\ \nonumber &+ \frac{\Delta t^2}{272160}\left(1283\mathbf{a}_0^{(1)}-7659\mathbf{a}_{1/3}^{(1)} - 2421\mathbf{a}_{2/3}^{(1)} - 163\mathbf{a}_1^{(1)}\right)\\
\mathbf{v}_{2/3} &= \mathbf{v}_0 + \frac{\Delta t}{1701}\left(223\mathbf{a}_0 + 540\mathbf{a}_{1/3} + 351\mathbf{a}_{2/3} + 20\mathbf{a}_1\right) + \frac{\Delta t^2}{8505}\left(43\mathbf{a}_0^{(1)}-144\mathbf{a}_{1/3}^{(1)} -171\mathbf{a}_{2/3}^{(1)} - 8\mathbf{a}_1^{(1)}\right)\\
\mathbf{v}_{1} &= \mathbf{v}_0 + \frac{\Delta t}{244}\left(31\mathbf{a}_0 + 81\mathbf{a}_{1/3} + 81\mathbf{a}_{2/3} + 31 \mathbf{a}_1\right) + \frac{\Delta t^2}{3360}\left(19\mathbf{a}_0^{(1)}-27\mathbf{a}_{1/3}^{(1)} + 27\mathbf{a}_{2/3}^{(1)} - 19\mathbf{a}_1^{(1)}\right).
\end{align}

Similarly, the 4-point scheme including two derivatives of the acceleration at each point is
\begin{align}
\mathbf{v}_{1/3} &= \mathbf{v}_0 + \frac{\Delta t}{7185024}\left(912523\mathbf{a}_0 + 1921077\mathbf{a}_{1/3} -473931\mathbf{a}_{2/3} + 35339\mathbf{a}_1\right) 
\\ \nonumber &+ \frac{\Delta t^2}{35925120}\left(214943\mathbf{a}_0^{(1)}-287739\mathbf{a}_{1/3}^{(1)} + 287739\mathbf{a}_{2/3}^{(1)} - 17823\mathbf{a}_1^{(1)}\right)
\\ \nonumber &+ \frac{\Delta t^3}{107775360}\left(11369\mathbf{a}_0^{(2)}+199035\mathbf{a}_{1/3}^{(2)} - 67077\mathbf{a}_{2/3}^{(2)} + 1513\mathbf{a}_1^{(2)}\right),
\\
\mathbf{v}_{2/3} &= \mathbf{v}_0 + \frac{\Delta t}{56133}\left(7031\mathbf{a}_0 + 24462\mathbf{a}_{1/3} + 5751\mathbf{a}_{2/3} + 178\mathbf{a}_1\right) \\ \nonumber
&+ \frac{\Delta t^2}{280665}\left(1632\mathbf{a}_0^{(1)}+864\mathbf{a}_{1/3}^{(1)} -864\mathbf{a}_{2/3}^{(1)} - 92\mathbf{a}_1^{(1)}\right)
+ \frac{\Delta t^3}{841995}\left(85\mathbf{a}_0^{(2)}+1060\mathbf{a}_{1/3}^{(2)} -171\mathbf{a}_{2/3}^{(2)} - 8\mathbf{a}_1^{(2)}\right),
\\
\mathbf{v}_{1} &= \mathbf{v}_0 + \frac{\Delta t}{9856}\left(1283\mathbf{a}_0 + 3645\mathbf{a}_{1/3} + 3645\mathbf{a}_{2/3} + 1283 \mathbf{a}_1\right) \\ \nonumber &+\frac{\Delta t^2}{49280}\left(311\mathbf{a}_0^{(1)}-243\mathbf{a}_{1/3}^{(1)} + 243\mathbf{a}_{2/3}^{(1)} - 311\mathbf{a}_1^{(1)}\right)
+\frac{\Delta t^3}{147840}\left(17\mathbf{a}_0^{(2)}+243\mathbf{a}_{1/3}^{(2)} + 243\mathbf{a}_{2/3}^{(2)} + 17\mathbf{a}_1^{(2)}\right).
\end{align}

\subsection{Five-point Quadratures}
First, the integration scheme using five equally-spaced points and both the acceleration at each point and its first derivative is
\begin{align}
\mathbf{v}_{1/4} &= \mathbf{v}_0 + \frac{\Delta t}{17418240}\left(1539551\mathbf{a}_0 + 1429936\mathbf{a}_{1/4} + 711936\mathbf{a}_{2/4} + 613456\mathbf{a}_{3/4} + 59681\mathbf{a}_1\right) \\ \nonumber &+ \frac{\Delta t^2}{11612160}\left(26051\mathbf{a}_0^{(1)}-249656\mathbf{a}_{1/4}^{(1)} -183708\mathbf{a}_{2/4}^{(1)} -49720\mathbf{a}_{3/4}^{(1)} - 2237\mathbf{a}_{1}^{(1)}  \right),\\
\mathbf{v}_{2/4} &= \mathbf{v}_0 + \frac{\Delta t}{272160}\left(24463\mathbf{a}_0 + 52928\mathbf{a}_{1/4} + 44928\mathbf{a}_{2/4} + 12608\mathbf{a}_{3/4} + 1153\mathbf{a}_1\right) \\ \nonumber &+ \frac{\Delta t^2}{181440}\left(241\mathbf{a}_0^{(1)}-3040\mathbf{a}_{1/4}^{(1)} -4536\mathbf{a}_{2/4}^{(1)} -992\mathbf{a}_{3/4}^{(1)} - 43\mathbf{a}_{1}^{(1)}  \right),\\
\mathbf{v}_{3/4} &= \mathbf{v}_0 + \frac{\Delta t}{71680}\left(6501\mathbf{a}_0 + 14736\mathbf{a}_{1/4} + 20736\mathbf{a}_{2/4} + 11376\mathbf{a}_{3/4} + 411\mathbf{a}_1\right) \\ \nonumber &+ \frac{\Delta t^2}{143360}\left(339\mathbf{a}_0^{(1)}-2232\mathbf{a}_{1/4}^{(1)} -2268\mathbf{a}_{2/4}^{(1)} -1464\mathbf{a}_{3/4}^{(1)} - 45\mathbf{a}_{1}^{(1)}  \right),\\
\mathbf{v}_{1} &= \mathbf{v}_0 + \frac{\Delta t}{17010}\left(1601\mathbf{a}_0 + 4096\mathbf{a}_{1/4} + 5616\mathbf{a}_{2/4} + 4096\mathbf{a}_{3/4} + 1601\mathbf{a}_1\right) \\ \nonumber &+ \frac{\Delta t^2}{11340}\left(29\mathbf{a}_0^{(1)}-128\mathbf{a}_{1/4}^{(1)} + 128\mathbf{a}_{3/4}^{(1)} - 29\mathbf{a}_{1}^{(1)}  \right).
\end{align}

The 5-point scheme including two derivatives of the acceleration at each point is
\begin{align}
\nonumber \mathbf{v}_{1/4} &=  \mathbf{v}_0 + \frac{\Delta t}{239117598720}\left(21200023187\mathbf{a}_0 + 70647841792\mathbf{a}_{1/4} - 47723240448\mathbf{a}_{2/4} + 16205888512\mathbf{a}_{3/4} -551113363\mathbf{a}_1\right) \\  &+ \frac{\Delta t^2}{159411732480}\left(457963769\mathbf{a}_0^{(1)}+681473536\mathbf{a}_{1/4}^{(1)} + 1215937008\mathbf{a}_{2/4}^{(1)} - 1001720320\mathbf{a}_{3/4}^{(1)} - 25533817\mathbf{a}_{1}^{(1)}  \right)\\
 \nonumber &+ \frac{\Delta t^3}{318823464960}\left(10906367\mathbf{a}_0^{(2)} + 406999552\mathbf{a}_{1/4}^{(2)} - 478158336\mathbf{a}_{2/4}^{(2)} + 86752768\mathbf{a}_{3/4}^{(2)} - 977791\mathbf{a}_{1}^{(2)}  \right),\\
\mathbf{v}_{2/4} &= \mathbf{v}_0 + \frac{\Delta t}{934053120}\left(82429429\mathbf{a}_0 + 382790144\mathbf{a}_{1/4} -51757056\mathbf{a}_{2/4} + 55514624\mathbf{a}_{3/4} -1950581\mathbf{a}_1\right) \\ \nonumber &+ \frac{\Delta t^2}{622702080}\left(1772191\mathbf{a}_0^{(1)}+5861120\mathbf{a}_{1/4}^{(1)} + 555984\mathbf{a}_{2/4}^{(1)} - 3444992\mathbf{a}_{3/4}^{(1)} - 90527\mathbf{a}_{1}^{(1)}  \right)\\
 \nonumber &+ \frac{\Delta t^3}{1245404160}\left(42001\mathbf{a}_0^{(2)} + 1729664\mathbf{a}_{1/4}^{(2)} -1465344\mathbf{a}_{2/4}^{(2)} + 301952\mathbf{a}_{3/4}^{(2)} -3473\mathbf{a}_{1}^{(2)}  \right),\\
\mathbf{v}_{3/4} &= \mathbf{v}_0 + \frac{\Delta t}{328007680}\left(29017419\mathbf{a}_0 + 131687424\mathbf{a}_{1/4} + 29113344\mathbf{a}_{2/4} + 57007104\mathbf{a}_{3/4} -819531\mathbf{a}_1\right) \\ \nonumber &+ \frac{\Delta t^2}{656015360}\left(1876707\mathbf{a}_0^{(1)}+5681664\mathbf{a}_{1/4}^{(1)} + 5003856\mathbf{a}_{2/4}^{(1)} - 6999552\mathbf{a}_{3/4}^{(1)} + 112995\mathbf{a}_{1}^{(1)}  \right)\\
 \nonumber &+ \frac{\Delta t^3}{1312030720}\left(44613\mathbf{a}_0^{(2)} + 1783296\mathbf{a}_{1/4}^{(2)} -1119744\mathbf{a}_{2/4}^{(2)} + 465408\mathbf{a}_{3/4}^{(2)} -4293\mathbf{a}_{1}^{(2)}  \right),\\ 
 \mathbf{v}_{1} &= \mathbf{v}_0 + \frac{\Delta t}{7297290}\left(628741\mathbf{a}_0 + 3424256\mathbf{a}_{1/4} -808704\mathbf{a}_{2/4} + 3424256\mathbf{a}_{3/4} + 628741\mathbf{a}_1\right) \\ \nonumber &+ \frac{\Delta t^2}{2432430}\left(6569\mathbf{a}_0^{(1)}+36352\mathbf{a}_{1/4}^{(1)}  - 36352\mathbf{a}_{3/4}^{(1)} - 6569\mathbf{a}_{1}^{(1)}  \right)\\
 \nonumber &+ \frac{\Delta t^3}{9729720}\left(301\mathbf{a}_0^{(2)} + 15872\mathbf{a}_{1/4}^{(2)} -22896\mathbf{a}_{2/4}^{(2)} +15872\mathbf{a}_{3/4}^{(2)} +301\mathbf{a}_{1}^{(2)}  \right).
\end{align}

\subsection{Gaussian Quadratures} \label{app:gaussian}
Gaussian quadratures optimize node placement and use all the same number of additional derivatives at each node. In this work, I will only consider quadratures that include both the initial and final points on each interval, so that given a starting point $t_0$ and a stopping point $t_1=t_0+\Delta t$, the quadrature nodes can be defined as $t_{\xi_i}=t_0+\xi_i\Delta t$, where $\xi_i$ are points chosen optimally on the interval $[0,1]$. Because I require the first point to coincide with $t_0$ and the final to coincide with $t_0+\Delta t$, a quadrature consisting of $n+1$ points will always have $\xi_0=0$ and $\xi_n=1$. 

First, the 4-point 14th-order quadrature ($n=3, r=2$) tested in Figure \ref{fig:alltimesym} has
\begin{equation}
\xi_1 = 0.2856612203688185  ;~~~~ \xi_2 =0.7143387796311815.
\end{equation}
Then for example, $\mathbf{v}_{\xi_1}$, $\mathbf{v}_{\xi_2}$, and $\mathbf{v}_{1}$ can be given by 

\begin{align}\label{eq:gauss4}
\mathbf{v}_{\xi_1} - \mathbf{v}_0 = \Delta t \sum_{i=0}^3\mathfrak{g}^{\xi_1}_{i,0}\mathbf{a}_{\xi_i} + \Delta t^2 \sum_{i=0}^3\mathfrak{g}^{\xi_1}_{i,1}\mathbf{a}^{(1)}_i +\Delta t^3 \sum_{i=0}^3\mathfrak{g}^{\xi_1}_{i,2}\mathbf{a}^{(2)}_i\\
\mathbf{v}_{\xi_2} - \mathbf{v}_0 = \Delta t \sum_{i=0}^3\mathfrak{g}^{\xi_2}_{i,0}\mathbf{a}_{\xi_i} + \Delta t^2 \sum_{i=0}^3\mathfrak{g}^{\xi_2}_{i,1}\mathbf{a}^{(1)}_i +\Delta t^3 \sum_{i=0}^3\mathfrak{g}^{\xi_2}_{i,2}\mathbf{a}^{(2)}_i\\
\mathbf{v}_{\xi_3} - \mathbf{v}_0 = \Delta t \sum_{i=0}^3\mathfrak{g}^{\xi_3}_{i,0}\mathbf{a}_{\xi_i} + \Delta t^2 \sum_{i=0}^3\mathfrak{g}^{\xi_3}_{i,1}\mathbf{a}^{(1)}_i +\Delta t^3 \sum_{i=0}^3\mathfrak{g}^{\xi_3}_{i,2}\mathbf{a}^{(2)}_i
\end{align}
where
{\small
\begin{align} \nonumber
\mathfrak{g}^{\xi_1}_{0,0} \!&= 0.1149937554732505  , 
&\mathfrak{g}^{\xi_1}_{1,0} =&  0.1789346210042610, 
\!\!\!\!&\mathfrak{g}^{\xi_1}_{2,0} \!&=\!-0.01205376008003244
\!\!\!\!&\mathfrak{g}^{\xi_1}_{2,0} \!&=\!0.003786603971339454  \\ \nonumber
\mathfrak{g}^{\xi_1}_{0,1} \!&= 0.004943721565374436  , 
&\mathfrak{g}^{\xi_1}_{1,1} =&  -0.01093615597185940, 
\!\!\!\!&\mathfrak{g}^{\xi_1}_{2,1} \!&=\!0.0008576506106479335
\!\!\!\!&\mathfrak{g}^{\xi_1}_{2,1} \!&=\!-0.0003548677009728731 \\
\mathfrak{g}^{\xi_1}_{0,2} \!&= 8.022182449382128\times10^{-5}  , 
&\mathfrak{g}^{\xi_1}_{1,2} =&  0.0007209666465828754, 
\!\!\!\!&\mathfrak{g}^{\xi_1}_{2,2} \!&=\!-0.0001778954142071303
\!\!\!\!&\mathfrak{g}^{\xi_1}_{2,2} \!&=\!9.413340203546800\times10^{-6} \\ \nonumber
\mathfrak{g}^{\xi_2}_{0,0} \!&= 0.1001813021432683  , 
&\mathfrak{g}^{\xi_2}_{1,0} =&  0.4080858539654247, 
\!\!\!\!&\mathfrak{g}^{\xi_2}_{2,0} \!&=\!0.2170974728811313
\!\!\!\!&\mathfrak{g}^{\xi_2}_{2,0} \!&=\!-0.01102584935864278 \\ \nonumber
\mathfrak{g}^{\xi_2}_{0,1} \!&= 0.003616308295513116 , 
&\mathfrak{g}^{\xi_2}_{1,1} =&  0.005481404225439174, 
\!\!\!\!&\mathfrak{g}^{\xi_2}_{2,1} \!&=\!-0.01555990958665065
\!\!\!\!&\mathfrak{g}^{\xi_2}_{2,1} \!&=\!0.0009725455688884463 \\
\mathfrak{g}^{\xi_2}_{0,2} \!&= 4.667152439310981\times10^{-5}  , 
&\mathfrak{g}^{\xi_2}_{1,2} =&  0.001838715843722247, 
\!\!\!\!&\mathfrak{g}^{\xi_2}_{2,2} \!&=\!0.0009398537829322411
\!\!\!\!&\mathfrak{g}^{\xi_2}_{2,2} \!&=\!-2.413695989716466\times10^{-6} \\ \nonumber
\mathfrak{g}^{\xi_3}_{0,0} \!&= 0.1039679061146077  , 
&\mathfrak{g}^{\xi_3}_{1,0} =&  0.3960320938853923, 
\!\!\!\!&\mathfrak{g}^{\xi_3}_{2,0} \!&=\!0.3960320938853923
\!\!\!\!&\mathfrak{g}^{\xi_3}_{2,0} \!&=\!0.1039679061146077 \\ \nonumber
\mathfrak{g}^{\xi_3}_{0,1} \!&= 0.003971175996485989, 
&\mathfrak{g}^{\xi_3}_{1,1} =&  0.004623753614791240, 
\!\!\!\!&\mathfrak{g}^{\xi_3}_{2,1} \!&=\!-0.004623753614791240
\!\!\!\!&\mathfrak{g}^{\xi_3}_{2,1} \!&=\!-0.003971175996485989 \\ \nonumber
\mathfrak{g}^{\xi_3}_{0,2} \!&= 5.608486459665661\times10^{-5}  , 
&\mathfrak{g}^{\xi_3}_{1,2} =&  0.001660820429515116, 
\!\!\!\!&\mathfrak{g}^{\xi_3}_{2,2} \!&=\!0.001660820429515116
\!\!\!\!&\mathfrak{g}^{\xi_3}_{2,2} \!&=\!5.608486459665661\times10^{-5} \\
\end{align}
}

Then, the 5-point 18th-order quadrature  ($n=3, r=2$) tested in Figure \ref{fig:alltimesym} has
\begin{equation}
\xi_1 = 0.1827252983626181  ;~~~~ \xi_2 =0.5;~~~ \xi_3 = 0.8172747016373819.
\end{equation}

and is given by 

\begin{align}\label{eq:gauss5}
\mathbf{v}_{\xi_1} - \mathbf{v}_0 = \Delta t \sum_{i=0}^4\mathfrak{h}^{\xi_1}_{i,0}\mathbf{a}_{\xi_i} + \Delta t^2 \sum_{i=0}^4\mathfrak{h}^{\xi_1}_{i,1}\mathbf{a}^{(1)}_i +\Delta t^3 \sum_{i=0}^3\mathfrak{h}^{\xi_1}_{i,2}\mathbf{a}^{(2)}_i\\
\mathbf{v}_{\xi_2} - \mathbf{v}_0 = \Delta t \sum_{i=0}^4\mathfrak{h}^{\xi_2}_{i,0}\mathbf{a}_{\xi_i} + \Delta t^2 \sum_{i=0}^4\mathfrak{h}^{\xi_2}_{i,1}\mathbf{a}^{(1)}_i +\Delta t^3 \sum_{i=0}^3\mathfrak{h}^{\xi_2}_{i,2}\mathbf{a}^{(2)}_i\\
\mathbf{v}_{\xi_3} - \mathbf{v}_0 = \Delta t \sum_{i=0}^4\mathfrak{h}^{\xi_3}_{i,0}\mathbf{a}_{\xi_i} + \Delta t^2 \sum_{i=0}^4\mathfrak{h}^{\xi_3}_{i,1}\mathbf{a}^{(1)}_i +\Delta t^3 \sum_{i=0}^3\mathfrak{h}^{\xi_3}_{i,2}\mathbf{a}^{(2)}_i \\
\mathbf{v}_{\xi_4} - \mathbf{v}_0 = \Delta t \sum_{i=0}^4\mathfrak{h}^{\xi_4}_{i,0}\mathbf{a}_{\xi_i} + \Delta t^2 \sum_{i=0}^4\mathfrak{h}^{\xi_4}_{i,1}\mathbf{a}^{(1)}_i +\Delta t^3 \sum_{i=0}^3\mathfrak{h}^{\xi_4}_{i,2}\mathbf{a}^{(2)}_i
\end{align}

where 

\begin{align} \nonumber
\mathfrak{h}^{\xi_1}_{0,0} \!&= 0.07211829723593874  , 
&\mathfrak{h}^{\xi_1}_{1,0} =&  0.1157096467963714, 
&\mathfrak{h}^{\xi_1}_{2,0} \!&=\!-0.007373675187543635, \\ \nonumber
\mathfrak{h}^{\xi_1}_{3,0} \!&= 0.003652244987831095  , 
&\mathfrak{h}^{\xi_1}_{4,0} =&  -0.001381215469979492, \\ \nonumber
\mathfrak{h}^{\xi_1}_{0,1} \!&= 0.001930096252931788  , 
&\mathfrak{h}^{\xi_1}_{1,1} =&  -0.004604103232273663, 
&\mathfrak{h}^{\xi_1}_{2,1} \!&=\!0.0003711730404611596, \\ \nonumber
\mathfrak{h}^{\xi_1}_{3,1} \!&=-0.0001444484758551239 , 
&\mathfrak{h}^{\xi_1}_{4,1} =&  8.163564230465385\times10^{-5}, \\ \nonumber
\mathfrak{h}^{\xi_1}_{0,2} \!&=1.929582964046563\times10^{-5}  , 
&\mathfrak{h}^{\xi_1}_{1,2} =&  0.0002035430418924740, 
&\mathfrak{h}^{\xi_1}_{2,2} \!&=\!-7.751104480707071\times10^{-5}, \\
\mathfrak{h}^{\xi_1}_{3,2} \!&= 2.585742448631765\times10^{-5}  , 
&\mathfrak{h}^{\xi_1}_{4,2} =& -1.352668222884570\times10^{-6}, \\ \nonumber
\mathfrak{h}^{\xi_2}_{0,0} \!&= 0.06081244176394390  , 
&\mathfrak{h}^{\xi_2}_{1,0} =&  0.2782729393755240, 
&\mathfrak{h}^{\xi_2}_{2,0} \!&=\! 0.1689426102314692, \\ \nonumber
\mathfrak{h}^{\xi_2}_{3,0} \!&= -0.01175123763514402 , 
&\mathfrak{h}^{\xi_2}_{4,0} =&  0.003723246264206928, \\ \nonumber
\mathfrak{h}^{\xi_2}_{0,1} \!&= 0.001300585030047832  , 
&\mathfrak{h}^{\xi_2}_{1,1} =&  0.003441368107937227, 
&\mathfrak{h}^{\xi_2}_{2,1} \!&=\!-0.009556347502715723,\\ \nonumber
\mathfrak{h}^{\xi_2}_{3,1} \!&=0.0005926568666708526 , 
&\mathfrak{h}^{\xi_2}_{4,1} =& -0.0002163305233847670, \\ \nonumber
\mathfrak{h}^{\xi_2}_{0,2} \!&=9.523007297273199\times10^{-6}  , 
&\mathfrak{h}^{\xi_2}_{1,2} =& 0.0005735382026565897, 
&\mathfrak{h}^{\xi_2}_{2,2} \!&=\!0.0005051828736679725, \\
\mathfrak{h}^{\xi_2}_{3,2} \!&= -7.757240858082574\times10^{-5}  , 
&\mathfrak{h}^{\xi_2}_{4,2} =& 3.513920049923857\times10^{-6}, \\ \nonumber
\mathfrak{h}^{\xi_3}_{0,0} \!&= 0.06591690349813032 , 
&\mathfrak{h}^{\xi_3}_{1,0} =&  0.2628694567525489, 
&\mathfrak{h}^{\xi_3}_{2,0} \!&=\! 0.3452588956504821, \\ \nonumber
\mathfrak{h}^{\xi_3}_{3,0} \!&=0.1508120549440086 , 
&\mathfrak{h}^{\xi_3}_{4,0} =& -0.007582609207787913, \\ \nonumber
\mathfrak{h}^{\xi_3}_{0,1} \!&= 0.001598551195737253  , 
&\mathfrak{h}^{\xi_3}_{1,1} =&  0.002704262765411251, 
&\mathfrak{h}^{\xi_3}_{2,1} \!&=\!0.0003711730404611596,\\ \nonumber
\mathfrak{h}^{\xi_3}_{3,1} \!&=-0.007452814473540038, 
&\mathfrak{h}^{\xi_3}_{4,1} =&0.0004131806994991888, \\ \nonumber
\mathfrak{h}^{\xi_3}_{0,2} \!&=1.438959557008163\times10^{-5}  , 
&\mathfrak{h}^{\xi_3}_{1,2} =&0.0004701083695894464, 
&\mathfrak{h}^{\xi_3}_{2,2} \!&=\!0.001087876792143016, \\
\mathfrak{h}^{\xi_3}_{3,2} \!&= 0.0002924227521832900, 
&\mathfrak{h}^{\xi_3}_{4,2} =& -6.258902293268579\times10^{-6}, \\ \nonumber
\mathfrak{h}^{\xi_4}_{0,0} \!&= 0.064535688028150824 , 
&\mathfrak{h}^{\xi_4}_{1,0} =&  0.26652170174037995, 
&\mathfrak{h}^{\xi_4}_{2,0} \!&=\! 0.33788522046293844,\\ \nonumber
\mathfrak{h}^{\xi_4}_{3,0} \!&= 0.26652170174037995 , 
&\mathfrak{h}^{\xi_4}_{4,0} =& 0.064535688028150824, \\ \nonumber
\mathfrak{h}^{\xi_4}_{0,1} \!&= 0.001516915553432599  , 
&\mathfrak{h}^{\xi_4}_{1,1} =&  0.002848711241266374, 
&\mathfrak{h}^{\xi_4}_{2,1} \!&=\!0\\ \nonumber
\mathfrak{h}^{\xi_4}_{3,1} \!&=-0.002848711241266374, 
&\mathfrak{h}^{\xi_4}_{4,1} =&-0.001516915553432599, \\ \nonumber
\mathfrak{h}^{\xi_4}_{0,2} \!&=1.303692734719706\times10^{-5}  , 
&\mathfrak{h}^{\xi_4}_{1,2} =&0.0004959657940757640, 
&\mathfrak{h}^{\xi_4}_{2,2} \!&=\!0.001010365747335945,\\
\mathfrak{h}^{\xi_4}_{3,2} \!&= 0.0004959657940757640 , 
&\mathfrak{h}^{\xi_4}_{4,2} =& 1.303692734719706\times10^{-5}. \\ \nonumber
\end{align}

 \bibliographystyle{elsarticle-harv} 
 \bibliography{references.bib}





\end{document}